\documentclass[iop,apj,useAMS,usenatbib]{emulateapj-rtx4}
\usepackage{amsmath,epsfig,url,natbib}
\usepackage{graphicx,epstopdf,float,color}
\usepackage{courier}
\usepackage[hang,flushmargin]{footmisc}
\usepackage{hyperref}
\hypersetup{colorlinks=true, citecolor=blue, filecolor=magenta, urlcolor=cyan, linkcolor=blue}
\def\lsim{\mathrel{\rlap{\lower4pt\hbox{\hskip1pt$\sim$}}
    \raise1pt\hbox{$<$}}}                
\def\gsim{\mathrel{\rlap{\lower4pt\hbox{\hskip1pt$\sim$}}
    \raise1pt\hbox{$>$}}}                

\submitted{}
\begin{document}
\title{{\it Herschel}$^*$ and {\it Hubble} study of a lensed massive dusty starbursting galaxy at $\lowercase{z}\sim3$}
\author{H. Nayyeri\altaffilmark{1}}
\author{A. Cooray\altaffilmark{1}}
\author{E. Jullo \altaffilmark{2}} 
\author{D. A. Riechers \altaffilmark{3}} 
\author{T. K. D. Leung \altaffilmark{3}} 
\author{D. T. Frayer \altaffilmark{4}} 
\author{M. A. Gurwell \altaffilmark{5}} 
\author{A. I. Harris \altaffilmark{6}} 
\author{R.\,J.~Ivison \altaffilmark{7,8}} 
\author{M. Negrello \altaffilmark{9}} 
\author{I. Oteo \altaffilmark{7,8}} 
\author{S. Amber \altaffilmark{10}} 
\author{A. J. Baker \altaffilmark{11}} 
\author{J. Calanog \altaffilmark{12}}
\author{C. M. Casey \altaffilmark{13}}
\author{H. Dannerbauer \altaffilmark{14,15}}
\author{G. De Zotti \altaffilmark{16}}
\author{S. Eales \altaffilmark{17}}
\author{H. Fu \altaffilmark{18}}
\author{M. J. Micha{\l}owski \altaffilmark{8}}
\author{N. Timmons \altaffilmark{1}}
\author{J. L. Wardlow \altaffilmark{19}}

\altaffiltext{$\star$}{{\it Herschel} is an ESA space observatory with
  science instruments provided by European-led Principal Investigator
  consortia and with important participation from NASA.}
\altaffiltext{1}{Department of Physics and Astronomy, University of
  California Irvine, Irvine, CA}
\altaffiltext{2}{Laboratoire d’Astrophysique de Marseille, Pôle de l’Étoile Site de Château-Gombert
38, rue Frédéric Joliot-Curie 13388 Marseille cedex 13 FRANCE}
\altaffiltext{3}{Department of Astronomy, Cornell University, Ithaca, NY, 14853}
\altaffiltext{4}{National Radio Astronomy Observatory, Green Bank, WV, 24944}
\altaffiltext{5}{Harvard-Smithsonian Center for Astrophysics, 60 Garden St., MS 42, Cambridge, MA 02138 USA }
\altaffiltext{6}{Department of Astronomy University of Maryland College Park, MD 20742}
\altaffiltext{7}{European Southern Observatory, Karl-Schwarzschild-Strasse 2, 85748 Garching, Germany}
\altaffiltext{8}{Institute for Astronomy, University of Edinburgh,
  Blackford Hill, Edinburgh EH9 3HJ, UK}
\altaffiltext{9}{School of Physics and Astronomy, Cardiff University,
  The Parade, Cardiff CF24 3AA, UK}
\altaffiltext{10}{Department of Physical Sciences, The Open University, Milton Keynes, MK7 6AA, UK}
\altaffiltext{11}{Department of Physics \& Astronomy, Rutgers, the State University of New Jersey, 136 Frelinghuysen Road, Piscataway, NJ 08854-8019}
\altaffiltext{12}{Department of Physical Sciences, San Diego Miramar
  College, San Diego CA, 92126}
\altaffiltext{13}{Department of Astronomy, University of Texas at
  Austin, RLM 16.218 2515 Speedway, Stop C1400, Austin, TX 78712-1205}
\altaffiltext{14}{Instituto de Astrofísica de Canarias (IAC), E-38205 La Laguna, Tenerife, Spain}
\altaffiltext{15}{Universidad de La Laguna, Dpto. Astrofísica, E-38206 La Laguna, Tenerife, Spain}
\altaffiltext{16}{INAF-Osservatorio Astronomico di Padova, I-35122
  Padova, Italy}
\altaffiltext{17}{School of Physics \& Astronomy, Cardiff University, Cardiff, UK}
\altaffiltext{18}{Department of Physics \& Astronomy, University of
  Iowa, Iowa City, Iowa 52242}
\altaffiltext{19}{Centre for Extragalactic Astronomy, Department of Physics, Durham University, South Road, Durham, DH1 3LE, UK}

\journalinfo{Accepted to the Astrophysical Journal}
\begin{abstract}
We present the results of combined deep Keck/NIRC2, {\it HST}/WFC3
near-infrared and {\it Herschel} far infrared observations of an
extremely star forming dusty lensed galaxy identified from the {\it
  Herschel} Astrophysical Terahertz Large Area
Survey ({\it H}-ATLAS J133542.9+300401). The galaxy is gravitationally
lensed by a massive WISE identified galaxy cluster at $z\sim1$. The
lensed galaxy is
spectroscopically confirmed at $z=2.685$ from detection of $\rm {CO (1 \rightarrow
  0)}$ by GBT and from detection of $\rm {CO (3 \rightarrow
  2)}$ obtained with CARMA. We use the combined spectroscopic and
imaging observations to construct a detailed model of the
background dusty lensed sub-millemter galaxy (SMG) which allows us to study the source plane properties
of the target. The best-fit lens model provide magnification of $\mu_{\rm
  star}=2.10\pm0.11$ and $\mu_{\rm dust}=2.02\pm0.06$ for the stellar
and dust components respectively. Multi-band data yields a magnification corrected star formation rate of
$1900(\pm200)\,M_{\odot}{\rm yr^{-1}}$ and stellar
mass of $6.8_{-2.7}^{+0.9}\times10^{11}\,M_{\odot}$ consistent with a
main sequence of star formation at $z\sim2.6$. The CO observations yield a molecular gas mass of
$8.3(\pm1.0)\times10^{10}\,M_{\odot}$, similar to the most massive
star-forming galaxies, which together with the high star-formation
efficiency are responsible for the intense observed star formation
rates. The lensed SMG has a very short gas depletion time scale of
$\sim40$\,Myr. The high stellar mass and small gas fractions observed
indicate that the lensed SMG likely has already formed most of its
stellar mass and could be a progenitor of the most massive elliptical
galaxies found in the local Universe. 

\end{abstract}


\section{Introduction}

Understanding the formation of galaxies and their subsequent evolution with cosmic time
are fundamental goals of observational astronomy. Galaxies are believed
to form in gas-rich environments \citep{Dekel2009} and assemble their
mass through constant gas accretion in secular evolutions
\citep{Dekel2009, Kruijssen2014, Narayanan2015}, mergers \citep{Kauffmann1993,
  Hopkins2008, Tacconi2008, Engel2010, Hopkins2013} or both \citep{Kormendy2004,
  Genzel2008, Barro2013}. The evolutionary track of
galaxies is accompanied by various phases of star formation which
could be triggered by events such as mergers and sustained by
processes such as gas accretion onto the potential wells of the
underlying dark matter halos \citep{Cole1994, Granato2004, Bower2006,
  Furlong2015}. Studying the physical processes responsible for
regulating star-formation is crucial in getting a better
understanding of galaxy formation and evolution \citep{Law2009,
  Hemmati2014, Hemmati2015}.

One of the main sites of star formation in the Universe at high
redshifts are sub-millimeter galaxies (for a recent review see:
\citealp{Casey2014}). These system are rich in gas
and dust and have measured star formation rates in
excess of hundreds to a few thousand solar masses per year \citep{Greve2005, Capak2008,
  Magnelli2012, Michalowski2016}. These dusty galaxies are readily
identified and studied in extra-galactic surveys at long wavelengths
\citep{Blain1999, Elbaz2011, Leroy2013, Scoville2016, Hemmati2017} at which they are most
luminous as the UV light emitted by the hot young stars
(produced by intense star-formation activity) is absorbed and re-radiated by dust.

Recent studies indicate a very rapid mass assembly and short
duty cycles (starburst phase) for the high redshift SMGs \citep {Greve2005, Tacconi2006, Tacconi2008,
  Riechers2011, Magnelli2012, Toft2014}
with time scales as short as $\rm \sim 100\,Myr$. The high
star formation rate is responsible for mass build-up in SMGs. In
fact, cosmological hydrodynamic simulations show that the SMGs are on
average very massive and reside in $\sim 10^{13}\,M_{\odot}$
halos at $z\sim2$ \citep{Dave2010}. This agrees with multi-band observations of
sub-millimeter samples of galaxies at similar redshifts with stellar
masses as high as few times $10^{11}\,M_{\odot}$ and median of
$7\times10^{10}\,M_{\odot}$ \citet{Hainline2011}. The rapid mass assembly in SMGs
over short time scales followed by gas reservoir depletion due to
intense star-formation could possibly explain the origin
of the most massive quiescent galaxies. In fact recent studies have shown
that the very high redshift SMGs could indeed be
the progenitors of the most massive quiescent systems \citep
{Nayyeri2014, Toft2014}. A better
understanding of the mass assembly and the underlying star formation
responsible for it is achieved with the knowledge of the full
spectral energy distribution (SED) of the galaxy. Given the amounts of
dust, these systems are intrinsically
faint at shorter wavelengths and
become bright at longer wavelengths. Even with intrinsic luminosities
of $\rm \sim 10^{13}\,L_{\odot}$ identifications of sub-millimeter galaxies at high redshifts are still
challenging. Gravitational lensing provides a unique tool to study this
obscured population of galaxies at high redshift. The signal boost
provided by gravitational lensing and the increase in spatial
resolution combined with robust lens modeling 
allow us to study the star forming regions within these galaxies at
sub-kpc scales \citep {Swinbank2010, Dye2015, Swinbank2015,
  Rybak2015}. In fact this tool has been successfully utilized in several recent works to study the physical
properties of gas-rich star forming systems at high redshift \citep
{Ivison1998, Ivison2000, Frayer1998, Frayer1999,
  Riechers2011b, Gavazzi2011, Swinbank2011, Bussmann2012, Fu2012,
  Fu2013, Bussmann2013, Rawle2014, Messias2014, Timmons2015,
Bussmann2015}.

Wide-area far-infrared surveys have been very successful
in detecting lensed sub-millimeter galaxies. The steep number counts and
the negative $k$-correction at sub-mm wavelengths give rise to a
high magnification bias such that the fraction of lensed sources brighter
than a given threshold is significantly larger than other
wavelengths \citep{Blain1996}. This has been used by several recent studies
\citep {Negrello2010, Wardlow2013, Nayyeri2016, Negrello2016} to identify samples of high
redshift lensed SMGs from {\it Herschel} observations as well as
populations of lensed SMGs from wide-field observations
using Planck \citep{Canameras2015} and in the mm band from South Pole
Telescope \citep{Mocanu2013, Weiss2013, Mancuso2016, Strandet2016} and
Atacama Cosmology Telescope \citep{Su2017}.

In this work we study the physical properties of a {\it Herschel}
identified sub-millimeter galaxy at $z=2.685$ that is lensed by a
foreground massive cluster at $z \sim 0.98$. The signal boost provided by the
gravitational lensing in combination with our high resolution deep imaging
and spectroscopy in the near and far infrared gives us a unique opportunity to study the star
formation activity and physical properties of this lensed system as a
very massive SMG candidate identified during peak epoch of
star-formation activity \citep{Madau2014}.

This paper is organized as follows. In Section 2 we describe our
photometric and spectroscopic observations of NA.v1.489. In Section 3 we discuss
our lens modeling of the system using high resolution photometric and
spectroscopic data of the foreground cluster system. We study the physical
properties of the lensed system in Section 4. The results
of the combined lens modeling, SED fitting and spectral line analysis
are discussed in Section 5 and in Section 6 we summarize our main
findings. Throughout this paper we assume a standard cosmology with
$H_0=70\,\text{kms}^{-1}\text{Mpc}^{-1}$, $\Omega_m=0.3$ and
$\Omega_\Lambda=0.7$. All magnitudes are in the AB system where
$\text{m}_{\rm AB}=23.9-2.5\times\text{log}(f_{\nu}/1\mu \text{Jy})$ \citep
{Oke1983}. 

\begin{figure*}[t]
\centering
\includegraphics[trim=1.5cm 1cm 0cm 1.5cm, scale=0.9]{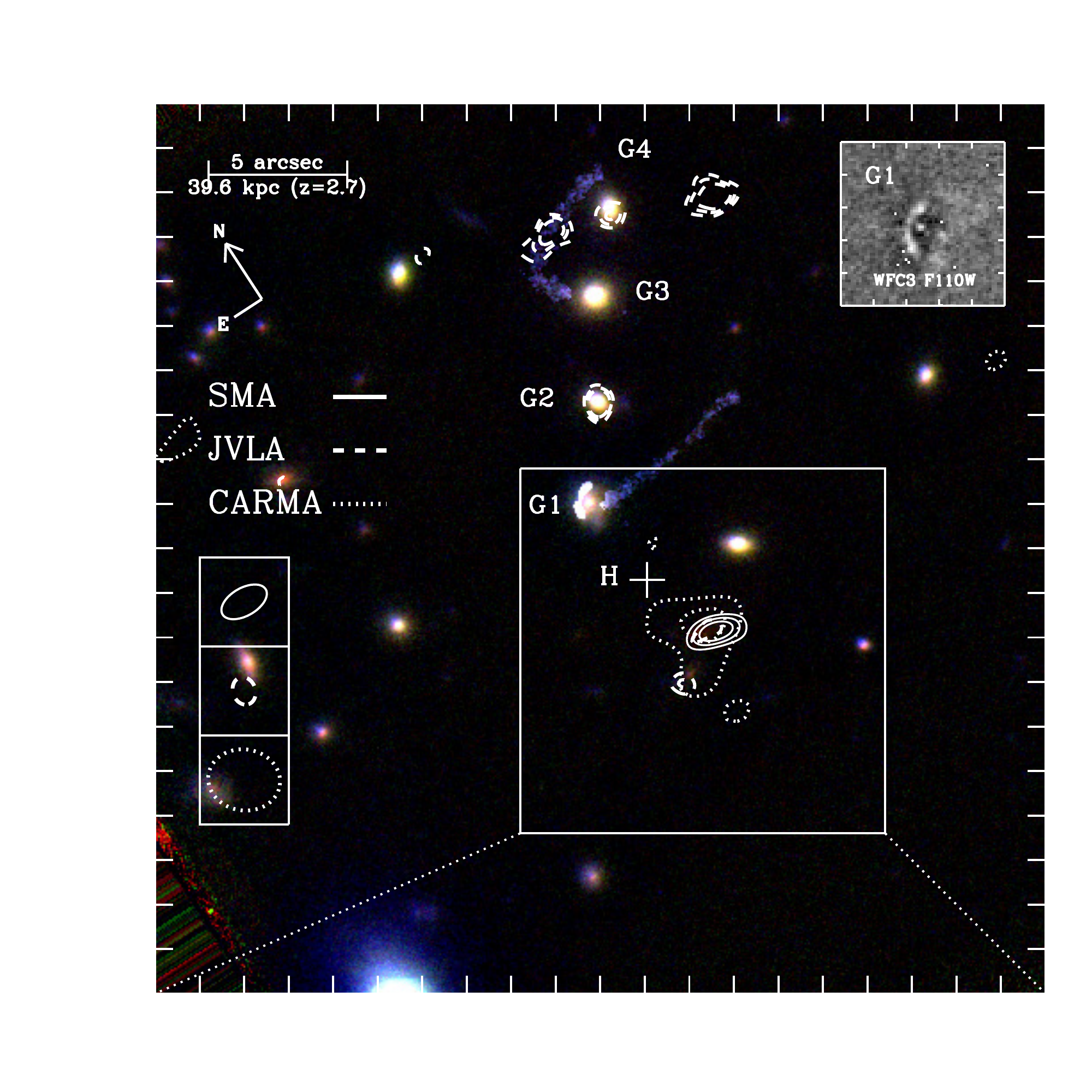} 
\includegraphics[trim=1.5cm 1.5cm 0cm 16.5cm, scale=0.9]{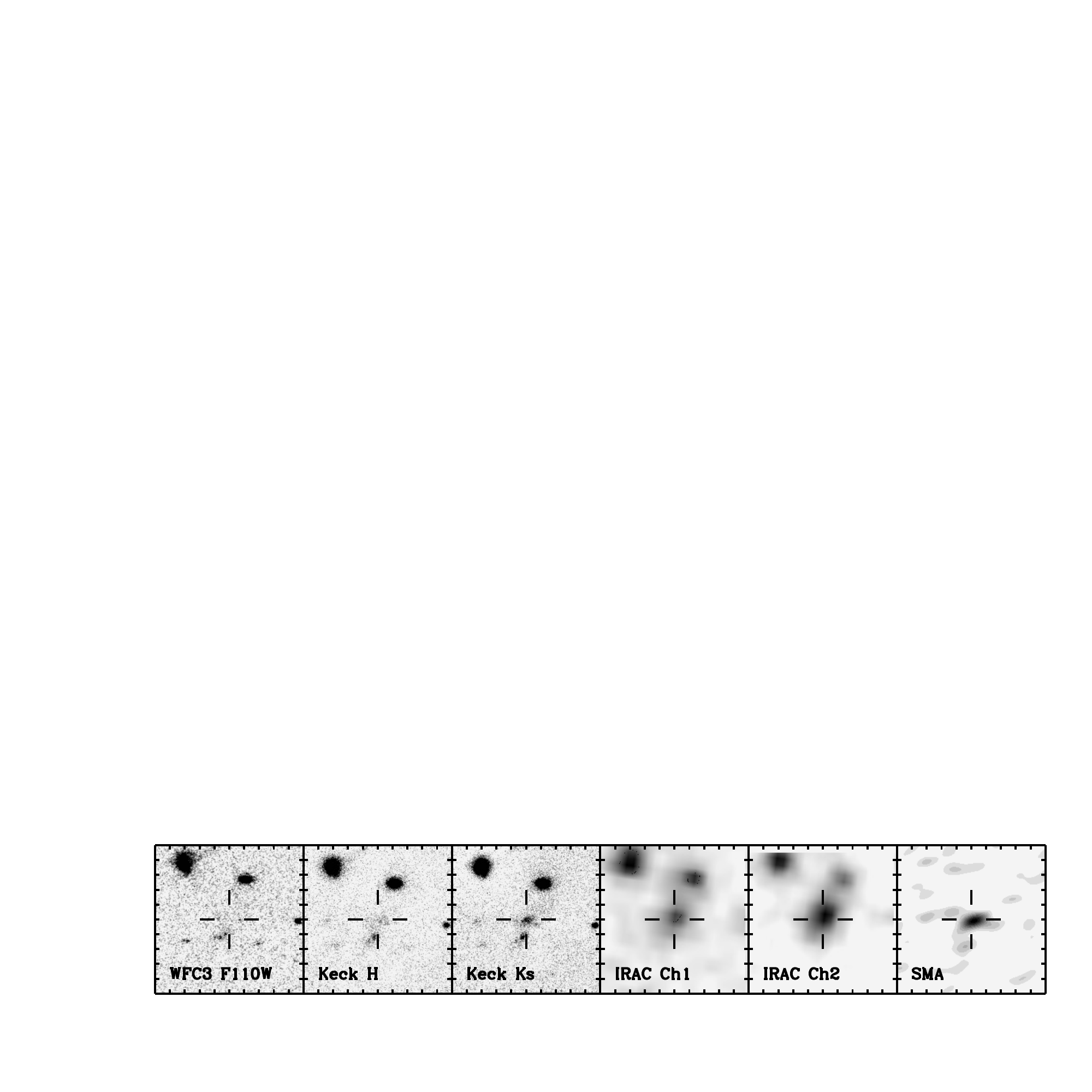} 
\caption{The {\it HST} F110W + Keck NIRC2 $H$ and $K_s$ band three color image of
  the NA.v1.489 system along with the four main deflecting galaxy
  potentials marked G1-G4 which are part of a foreground cluster at
  $z\sim0.98$ \citep{Stanford2014}. The solid box center marks the
  position of the high-$z$ SMG studied here and presented in the
  sub-panel. The JVLA, SMA and CARMA observations are shown with
  white contours on the image. The contours are drawn at 5$\times$,
  7$\times$ and 9$\times \sigma$ levels for JVLA and SMA and at 3$\times$,
  5$\times$ and 7$\times \sigma$ levels for CARMA ($\sigma_{\rm
    JVLA}=7.8\,\mu{\rm Jy/beam}$, $\sigma_{\rm
    SMA}=0.36\,{\rm mJy/beam}$, $\sigma_{\rm
    CARMA}=0.68\,{\rm Jy/beam\,kms^{-1}}$). The {\it Herschel} centroid is marked
  with a plus and is consistent with the peak SMA, CARMA and JVLA
  emissions given the PSF FWHM size of {\it Herschel}/SPIRE at
  250\,$\mu$m ($\sim 18^{\prime\prime}$). The top right box shows the zoomed-in F110W image
  of G1 revealing a lensed system. This together with the extended
  blue arcs (none of which are part of the SMG under study) are used
  to construct the lens model of the cluster. The radio emissions
  around G4 are likely radio lobes associated with a Faranaroff-Riley Type II
  (FR-II) radio source \citep{Fanaroff1974} which is also detected in
  the VLA FIRST \citep{Becker1994} and NVSS \citep{Condon1998}.}
\label{fig:Fig1}
\end{figure*}

\section{Data}

\subsection{Herschel Far-infrared Imaging}

{\it H}-ATLAS J133542.9+300401 lensed SMG (hereafter referred to as
NA.v1.489) was discovered by the {\it Herschel} Space Observatory
\citep{Pilbratt2010} as part of the Astrophysical Terahertz
Large Area Survey ({\it H}-ATLAS; \citealp{Eales2010}). The observations
were performed with the Spectral and Photometric Imaging
REceiver (SPIRE; \citealp{Griffin2010}) instrument at 250\,$\mu$m,
350\,$\mu$m and 500\,$\mu$m. NA.v1.489 was discovered as a SPIRE
500\,$\mu$m bright source (with $S_{500}>100$\,mJy) within {\it
  H}-ATLAS maps as a potential high-$z$ lensed SMG candidate
\citep{Negrello2016} with follow-up CO observations (described below)
revealing the high redshift nature of the source with $z=2.685$.

The {\it Herschel} images are processed with the {\it Herschel} Interactive
Processing Environment (HIPE; \citealp{Ott2010}) and are available
from the {\it Herschel} Science Archive. For the SPIRE photometry we used the point source catalog for
the {\it H}-ATLAS (\citealp{Valiante2016}, Maddox et al. in prep.)
which includes photometry in all the three SPIRE bands (at 250\,$\mu$m, 350\,$\mu$m
and 500\,$\mu$m). The PACS and SPIRE images will be released in future
studies (Smith et al. in prep.) along with the source catalogs (Maddox
et al. in prep.). The details of the object selection and photometry
is described in \citep{Valiante2016}.

\subsection{GBT Spectroscopy of $\rm {CO (1 \rightarrow 0)}$}

\begin{figure}
\centering
\includegraphics[trim=1cm 0cm 0cm 0cm, scale=0.44]{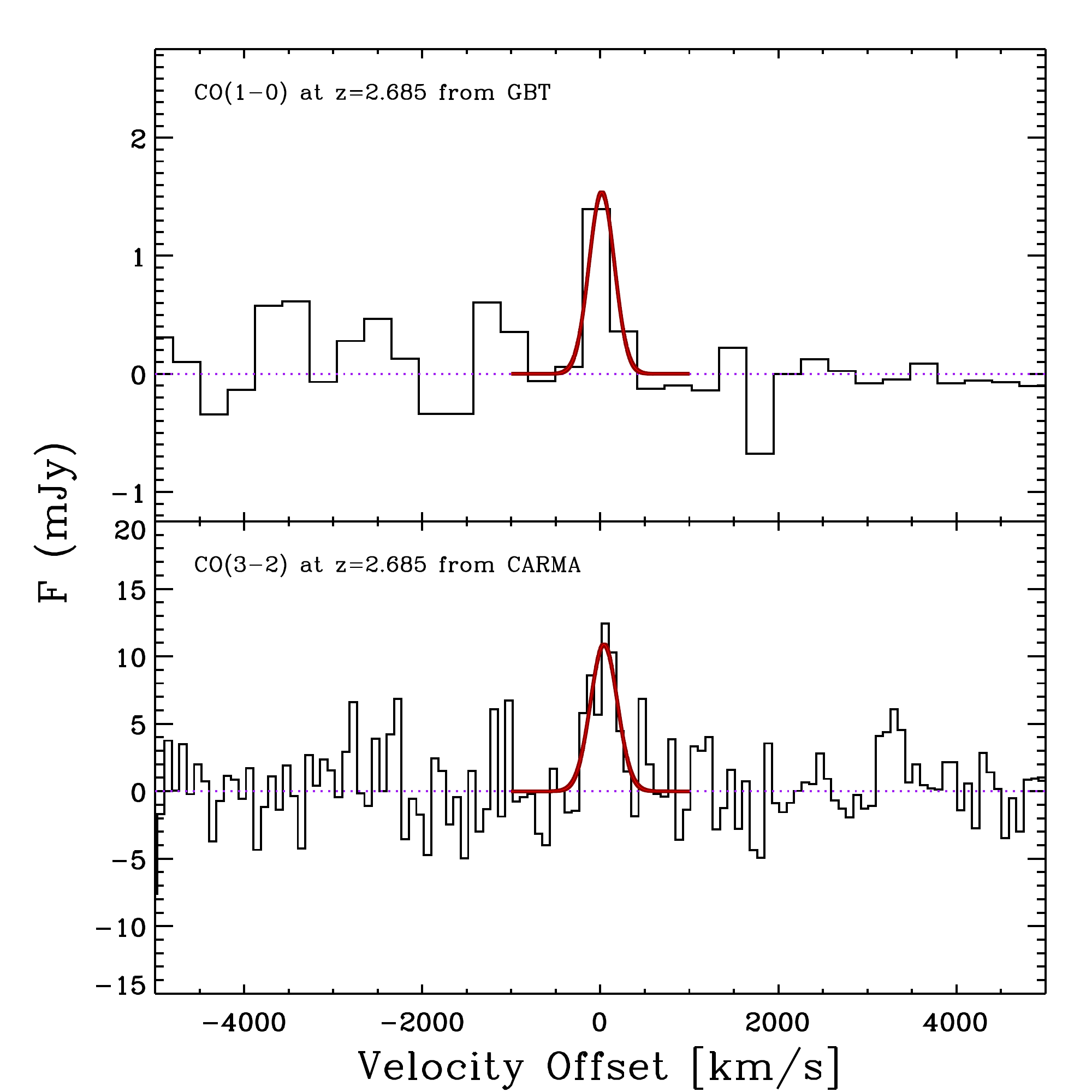}
\caption{The $\rm {CO (1 \rightarrow 0)}$ from GBT (top) and $\rm {CO
    (3 \rightarrow 2)}$ detection at $z = 2.685$ from CARMA (bottom)
  for the NA.v1.489. The red curves show the best-fit Gaussian to the
  lines centered on $v=0$ at $z=2.685$ for the 31.3\,GHz and 93.8\,GHz
  emission lines from GBT and CARMA. The best-fits have FWHM
  ($=2\sqrt{2{\rm ln}2}\,\sigma$) of  $305\pm87\,{\rm km/s}$ and
  $351\pm25\,{\rm km/s}$ and peak flux of 1.54\,mJy and 10.95\,mJy for the $\rm
  {CO (1 \rightarrow 0)}$ and $\rm {CO(3 \rightarrow
    2)}$ emissions respectively.}
\label{fig:Fig2}
\end{figure}

The NRAO\footnote{The National Radio Astronomy Observatory is a
facility of the National Science Foundation operated under cooperative
agreement by Associated Universities, Inc.} Green Bank Telescope (GBT)
was used to carry out the $\rm {CO (1 \rightarrow 0)}$ observations of NA.v1.489 during
three observing sessions (2012 November 29 and 30 and 2014 April 05; GBT
programs 12A299 and 13A137, PI: D. Frayer). These observations were
part of a comprehensive $\rm {CO (1 \rightarrow 0)}$ redshift survey
of {\it H}-ATLAS sources
using the Zpectrometer instrument on the GBT \citep{Frayer2011,
  Harris2012}. The observations were taken using the
sub-reflector beam switching (``SubBeamNod'') mode with a 10 second
switching interval.  Alternating sets of SubBeamNod observations
between the two targets were taken every 4 minutes to remove the
residual baseline structure. A total of 2.7\,hours of on-source
observations were obtained for NA.v1.489. The data were reduced
using the standard Zpectrometer data reduction package
\citep{Harris2012}. Based on the dispersion of measurements of the nearby
pointing source, we estimate a 15\% absolute calibration uncertainty
for the data. Figure~\ref{fig:Fig2} shows the GBT measured velocity
at $z=2.685$ (determined from combined GBT and CARMA observations; See
Figure~\ref{fig:Fig2} and Section below).


\subsection {CARMA Spectroscopy of $\rm {CO (3 \rightarrow 2)}$}

Observations of the $\rm {CO (3 \rightarrow 2)}$ rotational line
($\nu_{\rm rest}$ = 345.8\,GHz) towards the background galaxy
NA.v1.489 at $z = 2.685$ were carried out using the Combined Array for
Research in Millimeter-wave
Astronomy\footnote{\url{http://www.astro.caltech.edu/research/carma/}}
(CARMA) at the redshifted frequency of $\nu_{\rm obs}$ = 93.838\,GHz
(3.2\,mm; Program ID: cf0020, cf0025; P.I. Riechers).
Two observing runs were carried out on April 27 and June 9 2013 under excellent 3\,mm weather conditions in the C and D array configurations, respectively. The 3\,mm receivers were used to cover the redshifted $\rm {CO (3 \rightarrow 2)}$ line, employing a correlator setup providing a bandwidth of 3.75\,GHz in each sideband and spectral resolution of 5.208\,MHz ($\sim$17\,km\,s$^{-1}$). The line was placed in the
upper sidebands for both tracks with the local oscillator tuned to $\nu_{\rm LO}\sim$ 90.7\,GHz; this resulted in 1.9 hours and 2.6 hours of 15 antenna-equivalent on-source time after discarding unusable visibility data for C and D array observations, respectively.

For both tracks, the nearby radio quasar J1310+323 was observed every 15 minutes for
pointing, amplitude, and phase calibration, and MWC349 was observed as the primary
absolute flux calibrator. J1337$-$129 and 3C273 were observed as bandpass calibrators for C and D array observations, respectively, yielding $\sim
$15\% calibration accuracy.
The {\sc miriad} package was used to calibrate and analyze the visibility data which are imaged and de-convolved using
the {\sc clean} algorithm with ``natural'' weighting. This yields a synthesized clean beam size of 2$\farcs$6 $\times$ 2\farcs2 for the upper sideband image cube. The final rms noise is $\sigma$ = 0.68\,Jy\,km\,s$^{-1}$\,beam$^{-1}$ over a channel width of 208.3\,MHz (corresponding to 687\,km \,s$^{-1}$). 
The continuum image is created by averaging over all the line-free channels ($\nu_{\rm
  cont}\sim$90.7\,GHz). This yields a synthesized clean beam size of
3\farcs5 $\times$ 3\farcs0 and an rms noise of
0.24\,mJy\,beam$^{-1}$. These observations ultimately confirm the
redshift of NA.v1.489 to be $z = 2.685$, and thus, that the line
detected with the GBT indeed is $\rm {CO (1 \rightarrow 0)}$. They
also provide a precise position for the molecular gas reservoir, which
is consistent with that subsequently found for the dust emission based on SMA observations (see Section 2.8)

\subsection{Hubble Space Telescope WFC3 Imaging}

NA.v1.489 was observed with the {\it Hubble} Space Telescope Wide Field
Camera 3 ({\it HST}/WFC3) F110W filter (at 1.1\,$\mu$m) in Cycle 19 as part of
the SNAP imaging program of {\it Herschel} identified lensed SMGs
(PID: 12488; PI: Negrello; \citealp{Negrello2014}). The data were reduced using the {\sc iraf} MultiDrizzle
package with resampled pixel scale of 0.064$^{\prime\prime}$ using the adopted
dithering pattern \citep{Negrello2014}. The acquired image
has an exposure time of 252 seconds and reaches a 5$\sigma$ limiting
depth of 25.1 AB mag (over a $1^{\prime\prime}$
aperture). Figure~\ref{fig:Fig1} shows the {\it HST} image along with
additional observations. 

\subsection{Keck Adaptive Optics Imaging}

We observed NA.v1.489 in February 7 and 8, 2015 with the Keck/NIRC2
Adaptive Optics (AO) imaged (PID: U038N2L; PI: Cooray) in the $H$ and
$K_s$ band filters at 1.63\,$\mu$m and 2.15\,$\mu$m respectively with average seeing
of 0.6-0.7\,arcsec. The observations are done with a custom 9 point
dithering pattern for sky subtraction with 120 seconds and 80 seconds
exposures per frame at 0.04 arcsec/pixel. We
also acquired dark images with the hatch
closed and dome flats with and without the calibration lamps. The
individual frames are then co-added and flat and dark subtracted using
custom \texttt{IDL} routines. The combined images have exposure times
of 5640 seconds and 5280 seconds in the $H$ and $K_s$ bands, respectively. A natural guide
star of $R=17.3$ mag with a distance of 52.3\,arcsec was used for the
tip-tilt correction. Figure~\ref{fig:Fig1} shows the {\it HST} + Keck
combined image of the lensed system.


\subsection{Keck Optical Spectroscopy}

\begin{figure}
\centering
\leavevmode
\includegraphics[scale=0.4]{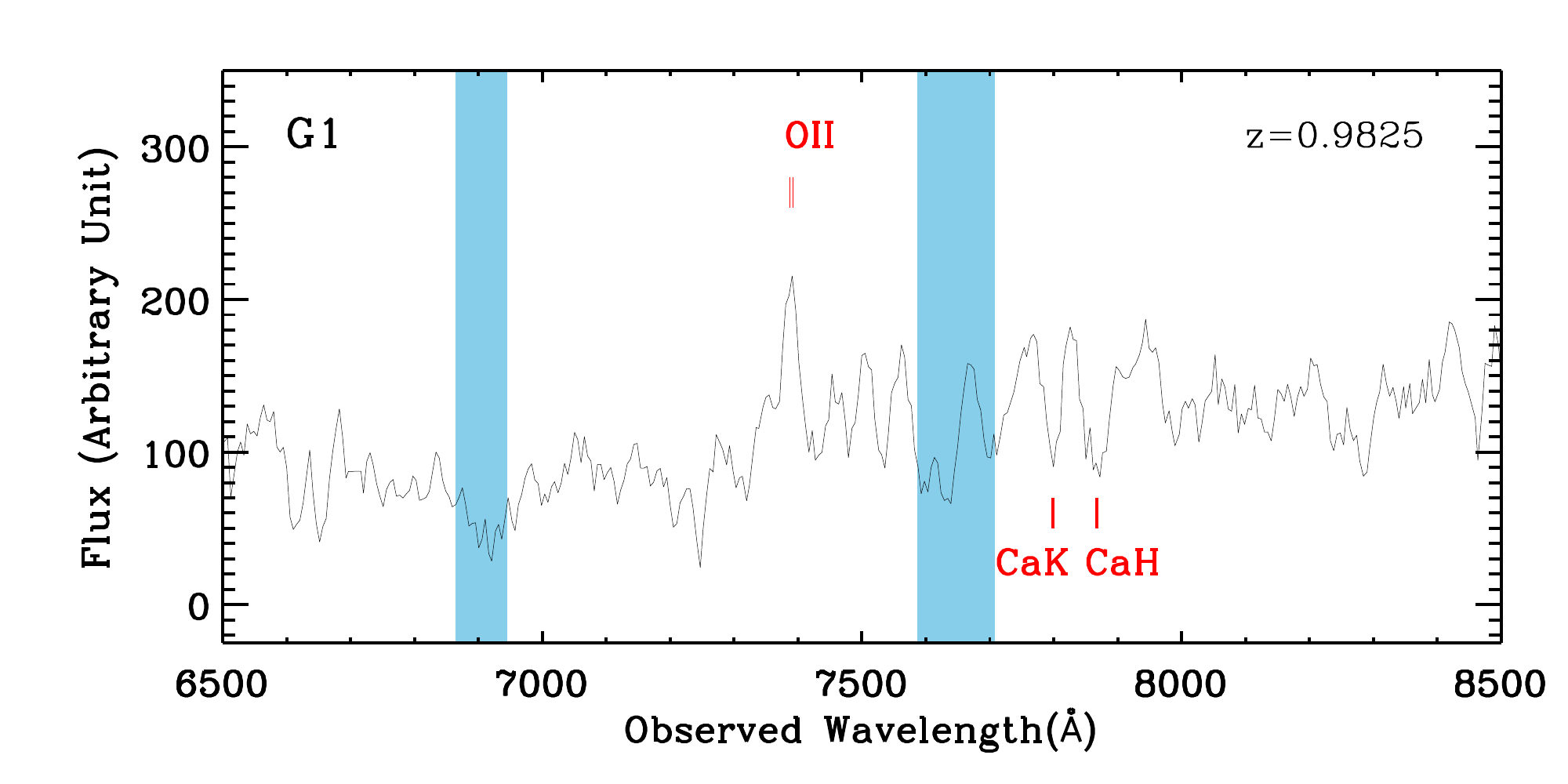} 
\includegraphics[scale=0.4]{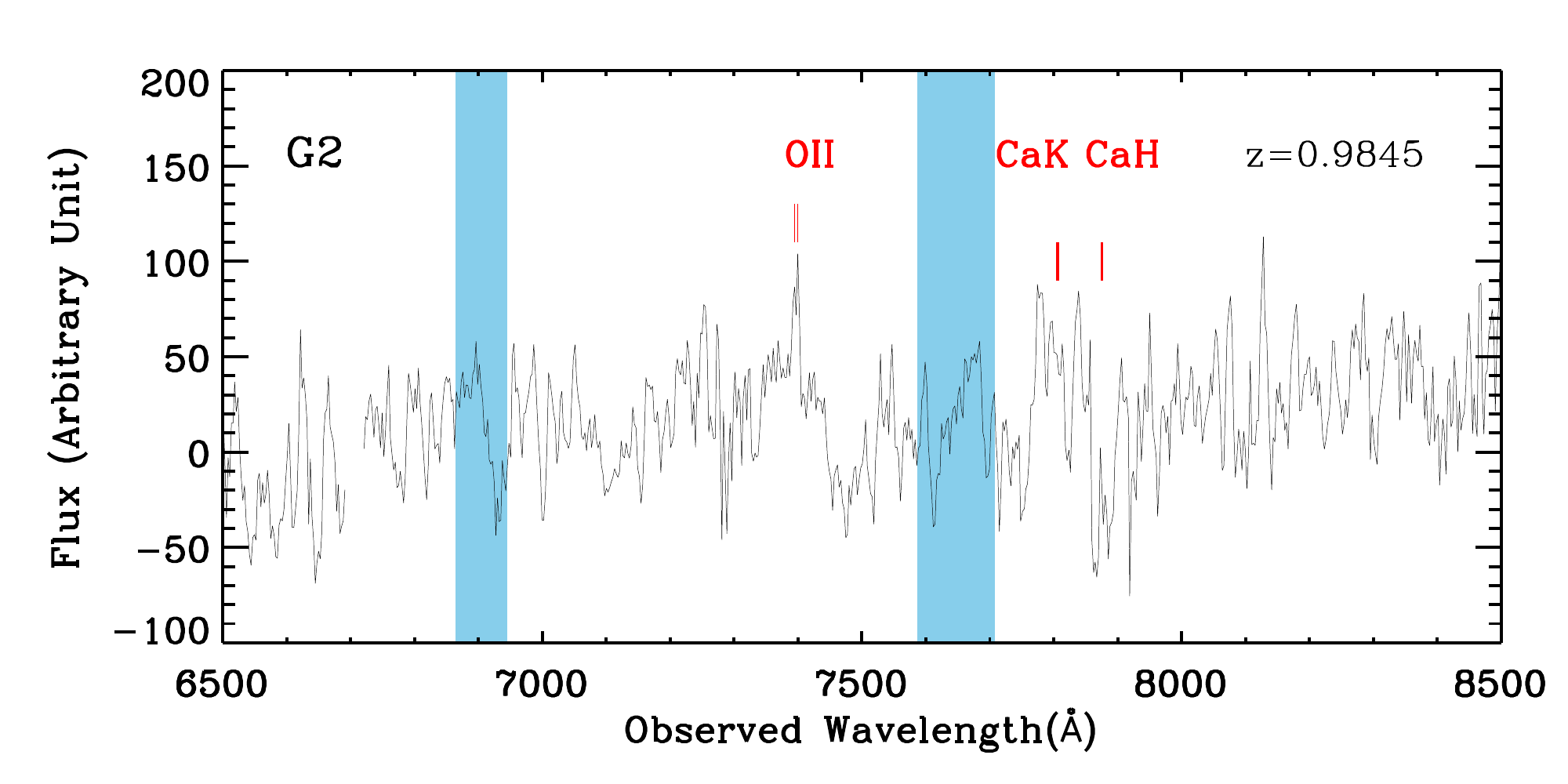} 
\includegraphics[scale=0.4]{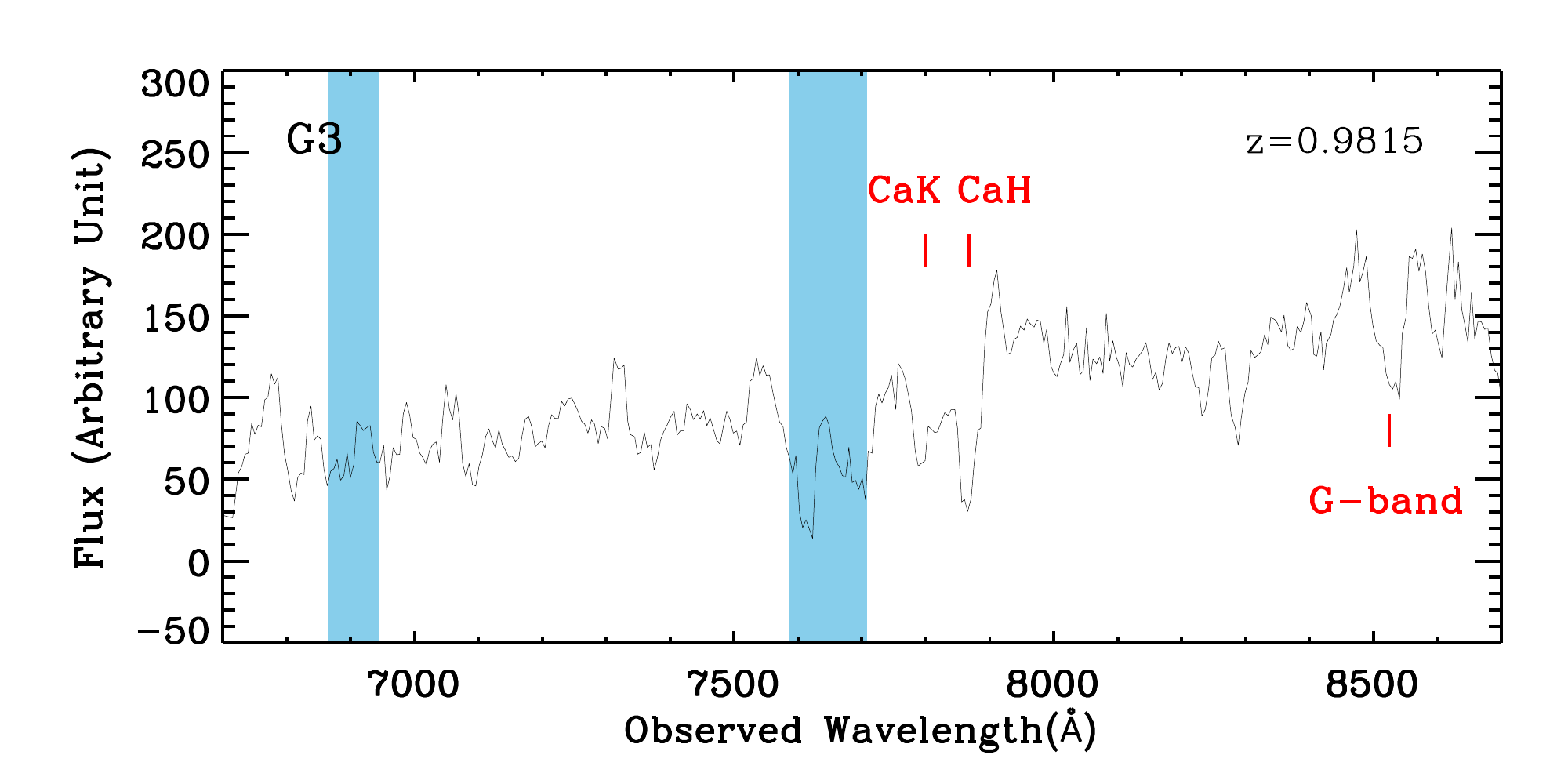} 
\includegraphics[scale=0.4]{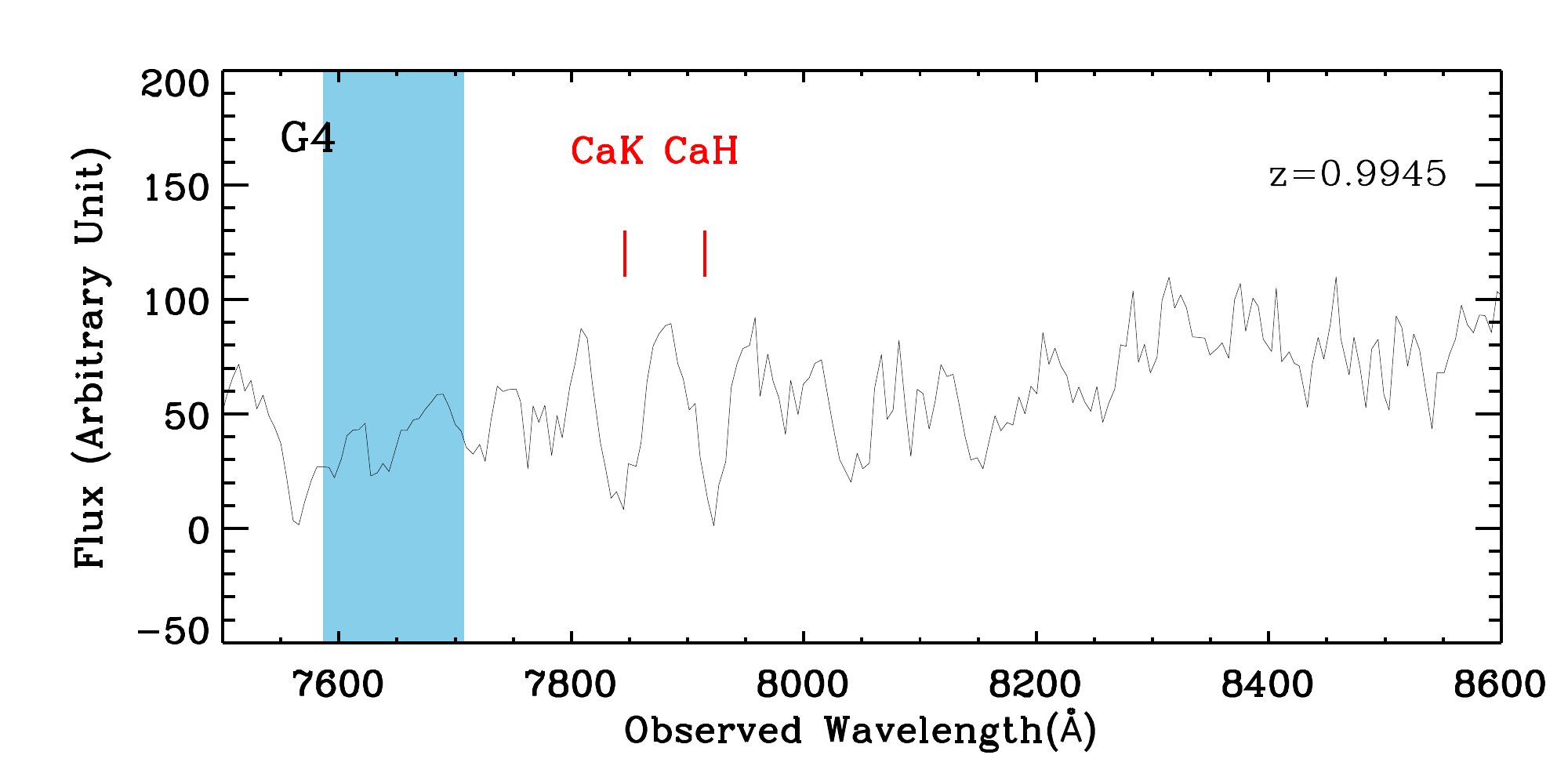} 
\caption{The Keck/DEIMOS 1D spectra of the foreground lensing galaxies
  extracted from the 2D spectra at the position of G1-G4
  (Figure~\ref{fig:Fig1}). The main atmospheric contamination in the
  wavelength range are marked as blue shaded regions. All the four
  galaxies have spectroscopic redshifts measured from the
  [O\,II]($\lambda3728, \lambda3729$) emission and/or
  Ca\,K($\lambda3934$) and Ca\,H($\lambda3968$) absorption putting them at
  $z=0.98-0.99$ consistent with the previous spectroscopic
  measurements for other cluster members in the field from GMOS on Gemini \citep{Stanford2014}.}
\label{fig:Fig3}
\end{figure}

The lensing system was observed with the Keck/DEIMOS optical
spectrograph \citep{Faber2003} in February 26, 2015 (PID:U029D; PI: Cooray). The G1-G4 foreground
galaxies (Figure~\ref{fig:Fig1}) were observed using the DEIMOS Long1.0B which is a single long slit of width 1 arcsec and
consists of 12 slitlets each 82 arcsec long. A PA of $-37^{\circ}$ was chosen to
align the galaxies on the slit with a star of $R=17.5$ mag at an offset
of 47$^{\prime\prime}$ and -29$^{\prime\prime}$ (east and north) with
respect to the center of the four lens used as
guiding. The foreground system was observed for
a total of 1100 seconds using the 600\,lines/mm grating with resolution of
3.5\AA\ and under clear conditions with $\sim$0.6 arcsec seeing. A central wavelength of 6700\AA\
blazed at 7500\AA\ was chosen for the observations giving a wavelength coverage of
${\rm 4050\AA - 9350\AA}$. 

The observed spectra where flat fielded and wavelength calibrated
using the {\sc deep2} pipeline. The 1D spectra were extracted at the
position of the four galaxies with optimal extraction \citep
{Horne1986}. Figure~\ref{fig:Fig3} shows the extracted 1D spectra for
the four co-aligned galaxies. We detect [O\,II] doublet $\lambda3728,
\lambda3729$ emission and/or Ca\,H\&K
absorption in the extracted 1D spectra putting the foreground system
at $z \sim 0.98$. This is the redshift that we adopt for our lens
modeling.

\subsection{JVLA Imaging}

The National Radio Astronomy Observatory’s (NRAO) JVLA\setcounter{footnote}{0}\footnote{This work is based on observations carried out with the
  JVLA. The NRAO is a facility of the NSF operated under cooperative
  agreement by Associated Universities, Inc} observations used in this
paper were carried out on February 15 and 17, 2015, when the array was
in it’s B-configuration (PI: R. Ivison, ID: 14B-475). Simultaneous 4\,GHz bandwidth in dual
polarization covering the $3.9-7.8$\,GHz range (C band). The
calibrator 3C286 was used for bandpass, phase, and flux calibration. The
data were calibrated and imaged in
{\sc casa}\footnote{\url{https://casa.nrao.edu/}} using standard calibration
techniques, including automatic RFI flagging. Image has
an rms of $\rm 7.8\,\mu Jy\,beam^{-1}$
with a synthesized beam size of $1.01^{\prime\prime} \times
0.81^{\prime\prime}$ at ${\rm PA = -71.6\,deg}$.

\subsection{Sub-Millimeter Array Imaging}

NA.v1.489 was initially observed using the Sub-Millimeter Array (SMA;
\citealp{Ho2004}) in December 2015. Observations
were performed on December 1 (2.6 hours on source, with 7 antennas) and December
7 (1.86 hours on source with 8 antennas) when the array was in its
most compact configuration. The
array was tuned to a local oscillator (LO) frequency of 228.3\,GHz
($\lambda=1.31$\,mm), and the integrated bandwidth was 6.25\,GHz per
sideband (12.5\,GHz total bandwidth). Both observations were obtained
in good weather, and emission was detected at $5.00\pm0.54$\,mJy, located
roughly 11$^{\prime\prime}$ from the phase center. 

Given the strong detection, further observations in the compact ($\sim
70$\,m max baseline for 2.8 hours on January 22, 2016) and extended ($\sim
220$\,m max baseline for 5.2 hours on April 14, 2016) SMA
configurations were obtained. The LO tuning was again 228.3\,GHz, with
$\sim 7$\,GHz bandwidth per sideband utilized (14\,GHz total
bandwidth). Phase and amplitude gain calibration was performed using
nearby sources J1310+323 and 3C286, with passband calibration
performed using 3C279, and the flux density scale set using Callisto,
known to 5\% at 1.3\,mm. Both observations were obtained under very good
weather (225\,GHz opacities of 0.07 in January, and 0.04 in April). The
observations were calibrated and ported into the NRAO Astronomy Image
Processing System ({\sc aips}). This is the data the we use in this
study.

Direct fitting of the visibility data determined the emission was best
fit by a Gaussian with a peak of $5.93\pm 0.43$\,mJy, offset from the
phase center by +1.9$^{\prime\prime}$ RA, -11.1$^{\prime\prime}$ Dec,
matching the integrated flux and
position determined at low resolution, and further suggested a
characteristic size of 0.9$^{\prime\prime}$ FWHM. The data were also imaged,
confirming again the positional offset and flux density. The
synthesized beam of the combined observations was $1.8^{\prime\prime}
\times 1.0^{\prime\prime}$.

\begin{figure*}[th]
\begin{center}
\includegraphics[trim=0cm 0cm 0cm 0cm,scale=0.28]{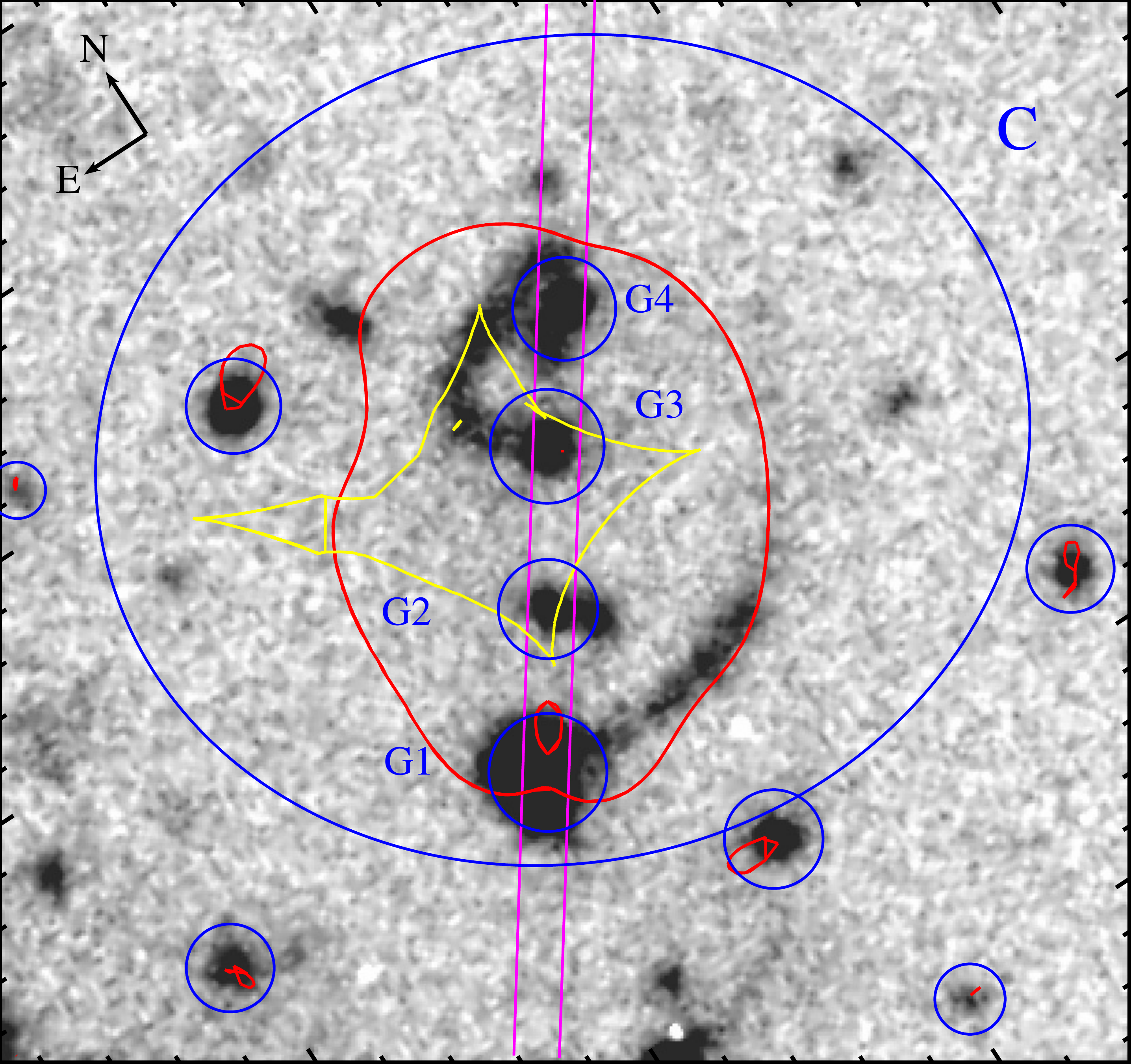}
\includegraphics[trim=-1cm 0cm 5cm 0cm,scale=0.28]{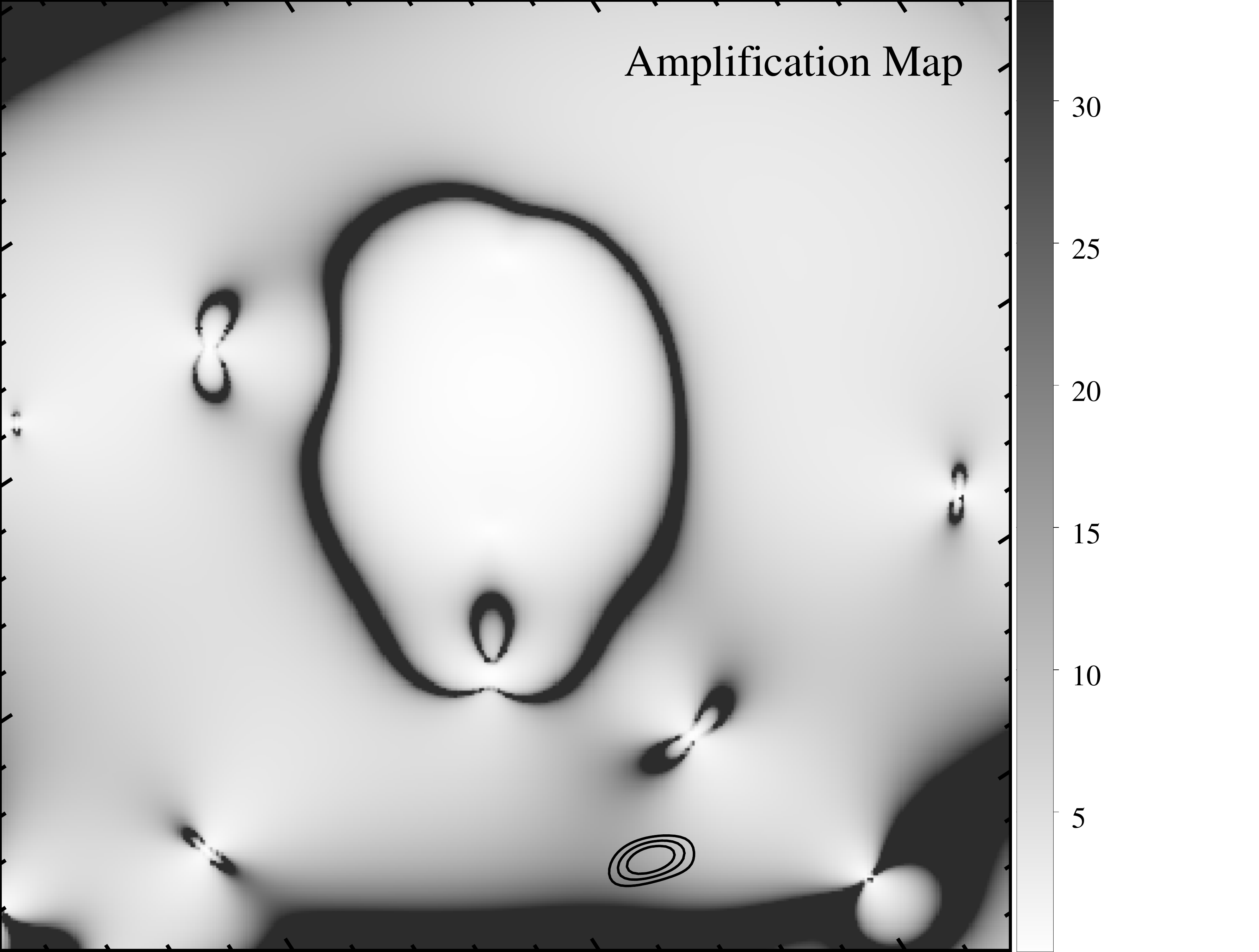}
\caption{{\it Left:} The Gemini r-band image of the foreground cluster lensing
  system. The blue lines show the foreground potentials at $z \sim 1$ used to construct the
lens model which include the main cluster potential (marked with C)
and the galaxy potentials, including the main four galaxies marked
G1-G4 (see Figure~\ref{fig:Fig1}), identified from spectroscopic observations by
Gemini \citep{Stanford2014} and our own Keck/DEIMOS observations (Keck
long-slit shown in magenta). The
red and yellow show the critical and caustic lines associated with the
best fit model. {\it Right:} The amplification map of the best-fit lens
model. The source magnification is estimated by averaging this map
across the object image area identified from the {\sc SExtractor} segmentation
maps. The black contours are from SMA
observations of NA.v1.489 (at 5$\times$, 7$\times$ and 9$\times
\sigma$ levels; see Figure~\ref{fig:Fig1}).}
\label{fig:Fig4}
\end{center}
\end{figure*}

\subsection {Archival Data}

The lensing cluster has been observed by the Gemini-North GMOS in both
imaging and spectroscopy modes (MOO J1335+3004; \citealp{Stanford2014}). The imaging
observations are in $r$ and $z$-bands (at 6300\AA\ and 9250\AA) with
total exposure times of 900 and 2160 seconds respectively
\citep{Stanford2014} and spectroscopic observations on candidates
identified from the optical catalogs with total
on-source exposure time of 6480 seconds. NA.v1.489 is not detected in these
Gemini observations. We use this spectroscopic and
imaging dataset to identify the foreground lensing galaxies and
distorted images associated with the lensing cluster as discussed in
the next Section. 

The foreground cluster was observed by the {\it Spitzer}/IRAC at
3.6\,$\mu$m and 4.5\,$\mu$m in Cycle 12 in warm mission (PI:
A. Gonzalez) and separately by the Wide-field Infrared Survey Explorer
(WISE; \citealp{Wright2010}) in all the four W1, W2, W3 and W4
bands. The IRAC observations were done with 30 seconds exposures in both
bands. The IRAC mosaics were in MJy/Sr which where converted to
$\mu$Jy/pixel using a pixel scale of 0.6\,arcsec. NA.v1.489 is
detected in both IRAC bands with at least ${\rm S/N}\sim9$.

\subsection {Multi-band Photometry}

We measured photometry of the background SMG using
{\sc SExtractor} \citep{Bertin1996}. The zero points for the Keck
observations were computed by comparing the photometry of bright
targets in the field to those in the UKIRT Infrared Deep Sky Survey (UKIDSS;
\citealp{Lawrence2007}). We run {\sc SExtractor} in dual mode with the Keck
$K_s$ image as the detection band for the {\it HST} and Keck photometry. 

For the IRAC observations we use each band as its own detection when
running {\sc SExtractor}. We measure the WISE flux density of NA.v1.489 through
aperture photometry with $r_{\rm ap}={\rm PSF FWHM}$ in each band
centered on the {\it Spitzer}/IRAC 3.6\,$\mu$m centroid. WISE aperture
photometry overestimates the fluxes in the W1 and W2 bands compared to
the IRAC 3.6\,$\mu$m and 4.5\,$\mu$m observations by $\sim 29\%$ due
to blending. We corrected the WISE W3 and W4 fluxes with the same
factor to account for blending.

{\it Herschel}/SPIRE photometry, as discussed above, is from the
{\it H}-ATLAS catalog of \citet{Valiante2016}. For the longer wavelength
SMA and JVLA, we first convert the maps from Jy/beam unit to mJy/pixel
given the beam size of the observations and the data pixel scale and
perform photometry on each image individually. Table 1 summarizes the
photometry extracted for the background SMG. We use this multi-band
photometry to construct the SED of the lensed galaxy in Section 4. 

\section{Lensing Model}

NA.v1.489 system is gravitationally lensed by a foreground cluster of
galaxies at $z=0.98$ first identified in the WISE survey by
\citet{Stanford2014} with follow-up spectroscopic and photometric
observations with Gemini as discussed above. The extended arcs
associated with lensing (of a galaxy at an unknown redshift) are visible in the high resolution HST
image of the cluster (blue arcs in Figure~\ref{fig:Fig1}). The SMG (marked in
Figure~\ref{fig:Fig1} with the solid box) was
independently identified from {\it Herschel} 500\,$\mu$m observations
by \citet{Negrello2016} as a potential lensed candidate. We performed lens modeling of this system
using the publicly available code of
{\sc lenstool}\footnote{\url{http://projets.lam.fr/projects/lenstool/wiki}}
\citep{Kneib1996, Jullo2007, Jullo2009}. {\sc lenstool} performs
Bayesian optimization given the redshift
and location of the identified images of the lensed sources in the
image plane. We used the two blue extended arcs identified in the
G1-G4 system (Figure~\ref{fig:Fig1}) along with the arc produced by G1 to constrain the
lens model. We used {\sc SExtractor} on the Gemini r-band to
identify the peak positions of the arcs and counter images in the
image plane as required by
{\sc lenstool}. We allowed the redshift of the arcs to vary during
the optimization. In lens modeling analysis we
used a combination of a cluster potential together with galaxy potentials as the
foreground deflecting components. We use a NFW profile
\citep{Navarro1996} for the lensing cluster at $z=0.98$ consistent
with the spectroscopic observations of various cluster members. For
the galaxy-scale potentials we used the spectroscopic identifications
from Gemini \citep{Stanford2014} and our Keck/DEIMOS observations as
the main foreground system combined with a red-sequence identified
cluster members from the color-magnitude diagram from the Gemini+{\it
  HST} extracted photometry. Figure~\ref{fig:Fig4} shows the
foreground potentials used to construct the lensing system. We
additionally keep the cluster potential
ellipticity and position variable, with the allowed centroid offset of
10 arcsec around the initial value (at the center of the four
galaxies). 

The output of the optimization process provides the best
estimate positions and properties of the deflecting potentials and is
considered the best-fit model. Figure~\ref{fig:Fig4} shows the
best-fit model computed by {\sc lenstool} along with the generated
amplification map. We use this map to measure the magnification of the
SMG at the SMA peak position. Averaging the amplification map across
the corresponding images identified from the {\sc SExtractor}
segmentation maps yield magnifications of $\mu_{\rm
  star}=2.10\pm0.11$ and $\mu_{\rm dust}=2.02\pm0.06$ for the stellar
and dust components respectively from modes of the computed Bayesian
model. We note here that the magnification for
both emissions (stellar and dust) are computed from the same
amplification map (constructed from the Gemini r-band) for the lensing
cluster due to lack of multiple images associated with dust that would
allow a separate lens modeling. In the next Section we de-magnify the
  stellar and dust continuum fluxes by the above-measured values
  before measuring the physical properties through SED fitting.

\begin{figure*}
\centering
\leavevmode
\includegraphics[trim=2cm 1.5cm 0cm 14.5cm,scale=0.99]{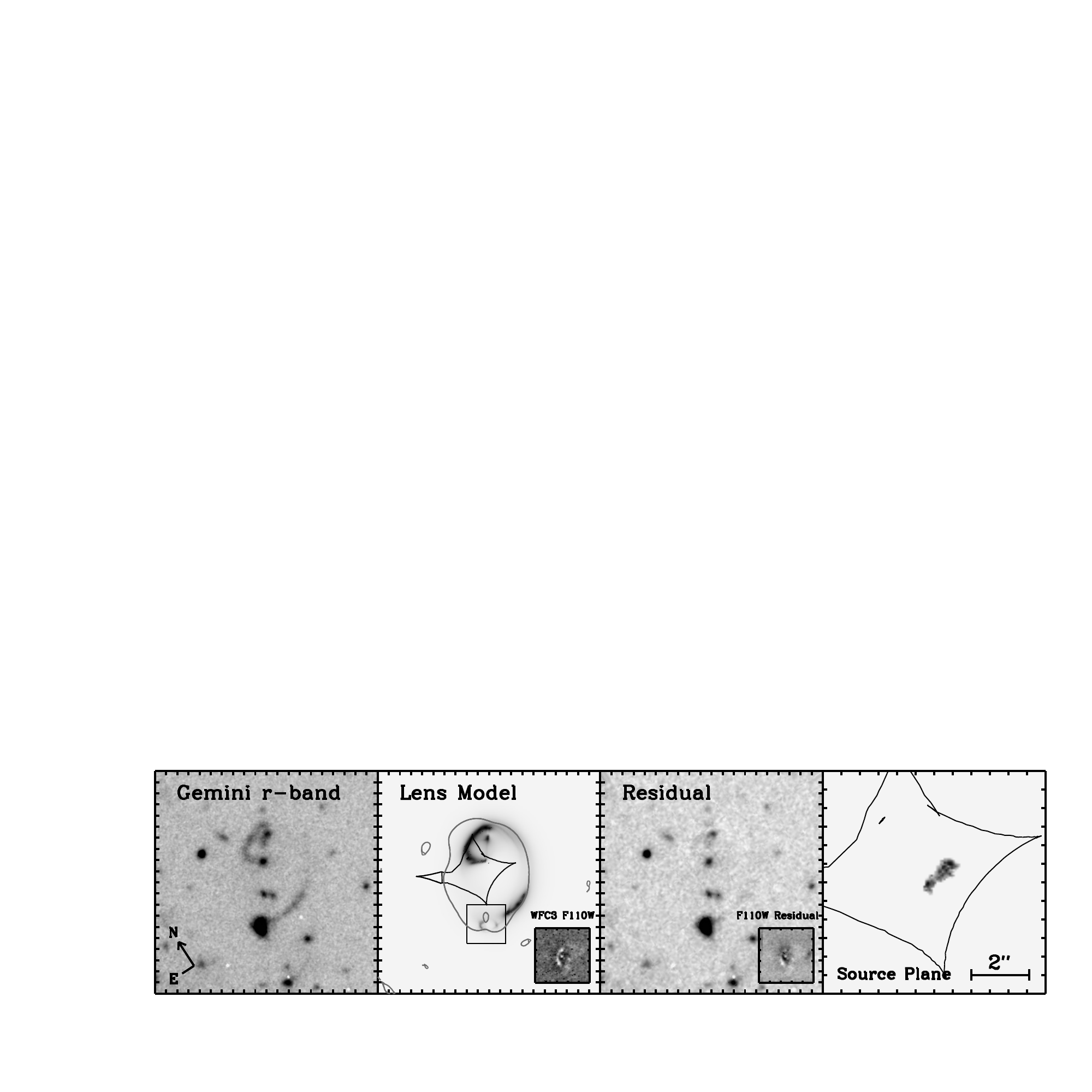}
\caption{The lens model of the foreground cluster constructed by
  {\sc lenstool} using the Gemini r-band image. The lens panel
  further shows the critical and caustic lines constructed by the
  model. The sub-panels in the lens model and residual maps represent
  the F110W image and residual of the small arc around G1
  respectively. The source plane reconstruction of the blue arcs (Figure~\ref{fig:Fig1}) in the
  cluster field used for building the model is shown in the far
  right.}
\label{fig:Fig5}
\end{figure*}

\begin{figure*}
\centering
\leavevmode
\includegraphics[trim=2cm 1.5cm 0cm 14.5cm,scale=0.99]{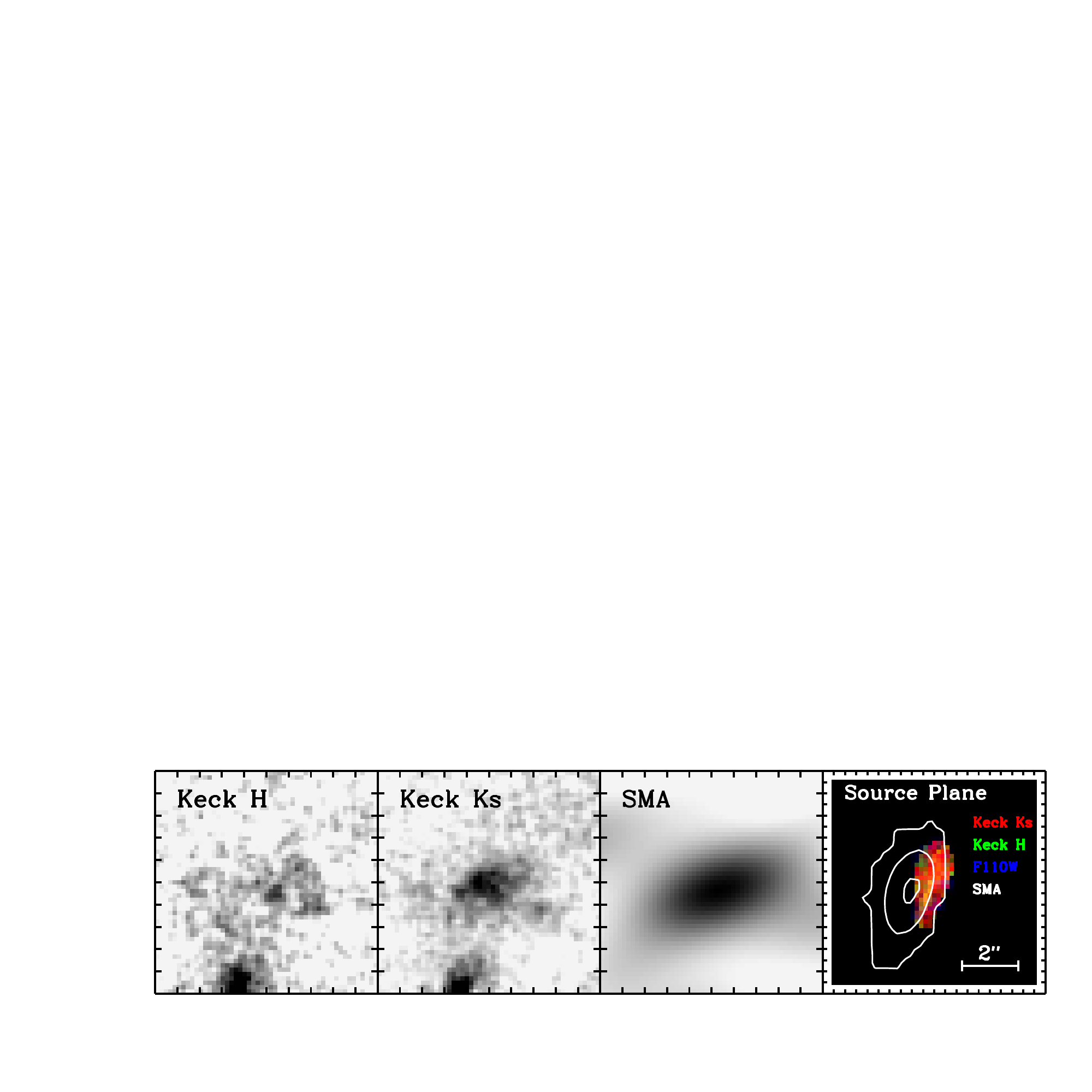}
\caption{The observed Keck $H$ and $K_s$ images of NA.v1.489 along
  with the SMA 1.31\,mm observations. The right panel shows the
  combined three-color source reconstructed image along with the SMA contours.}
\label{fig:Fig6}
\end{figure*}

Figure~\ref{fig:Fig5} shows the Gemini r-band image used in lens
modeling along with the re-constructed lens model in the image plane and
the residual map. The reconstructed image is generated by tracing the
source plane image through the best-fit lens model. We further show
the source plane reconstruction of the blue arcs used in generating
the model in this Figure for reference. Figure~\ref{fig:Fig6} shows
the Keck and SMA images of the lensed SMG NA.v1.489, along with the
source plane reconstructed three color image using the lens model
computed for the cluster as discussed above. The reconstructed source
plane has red colors as expected from the observed SED and consistent
with other rest-frame optical studies of SMGs. The rest-frame optical
emission, as arising from stellar light, has a co-moving distance
$\sim0.6\,$kpc from SMA dust emission in the
source plane. The stellar and dust emissions have half-light radii of
0.63\,kpc and 1.86\,kpc respectively measured from a S\'{e}rsic profile
using {\sc galfit} \citep{Peng2002}. This shows a more extended dust emission that is also
offset from the stellar light as reported previously in the literature
for high-$z$ SMGs \citep{Hodge2015, Spilker2015} and would
  suggest the presence of differential magnification (yet not
  significant) as reported above for the stellar and dust components. 

\section{Physical properties of NA.v1.489}

\subsection{SED inferred parameters}

We fit the SED of the background SMG with a library of model templates using the publicly
available SED fitting code of {\sc magphys} \citep{Cunha2008}. {\sc magphys} uses the
\citet{Bruzual2003} synthesis models for the stellar light and the
attenuation by \citet{Charlot2000} to compute the total infrared
luminosity absorbed and reradiated by dust using different dust
components at different wavelengths based on an energy balance scheme
\citep{Cunha2008}. We used an updated version of the {\sc magphys} code that
is better-suited for $z>1$ sub-millimeter galaxies \citep
{Cunha2015}. Briefly, it does so by extending the SED parameter priors to the high redshift,
high optical depth and actively star forming regime that is typical of
star forming galaxies at high redshift by adding new star formation
histories and dust attenuation recipes \citep{Cunha2015}.
 
\begin{figure}[t]
\centering
\leavevmode
\includegraphics[trim=1.5cm 0cm 0cm 0cm,scale=0.48]{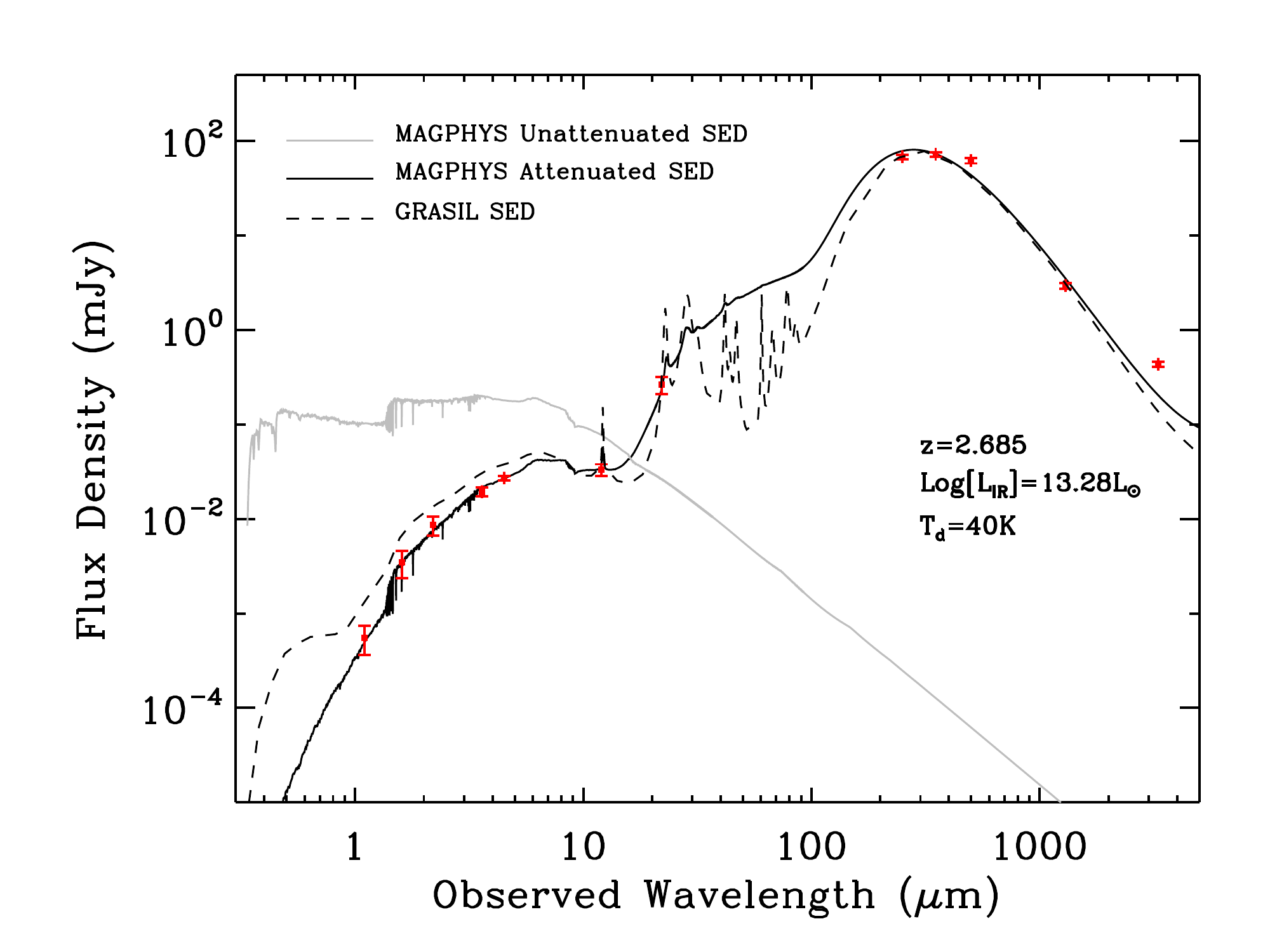}
\caption{The best-fit SED of the de-magnified flux density of NA.v1.489 using
  {\sc magphys} \citep{Cunha2008}. During the fitting process the redshift is fixed to
the spectroscopic redshift measured from CO observations. The fit uses
photometry from {\it HST}/WFC3 in the F110W band
  along with our Keck/NIRC2 observations in the $H$ and $K_s$ band. The
  infrared data is from {\it Spitzer} IRAC observations in the 3.6\,$\mu$m
  and 4.5\,$\mu$m and WISE W3 and W4 bands at 12.0\,$\mu$m and
  22.0\,$\mu$m respectively. The far-infrared data is from {\it Herschel}/SPIRE observations in
the 250\,$\mu$m, 350\,$\mu$m and 500\,$\mu$m bands from which the lensing system is
originally identified. We also use the SMA 1.31\,mm observations in
the SED fitting. These data
are shown with the red points on top of the best-fit SED (black line)
with the un-attenuated SED plotted in grey. The best-fit SED yields a
total infrared luminosity of $1.9\times10^{13}\,{L_{\odot}}$.}
\label{fig:Fig7}
\end{figure}

We use the photometry of NA.v1.489 outlined in Table 1 and correct
it for the magnification (reported in the previous section) as an input for SED
fitting in {\sc magphys}. This includes near-infrared data from {\it
  HST}/WFC3 F110W, Keck $H$ and $K_s$,
infrared measurements by {\it Spitzer}/IRAC in 3.6\,$\mu$m and
4.5\,$\mu$m, WISE observations at 12.0\,$\mu$m and 22.0\,$\mu$m,
far-infrared photometry from {\it Herschel}/SPIRE in the 250\,$\mu$m,
350\,$\mu$m and 500\,$\mu$m bands and 1.31\,mm observations by
SMA. For the SED fitting we fixed the redshift of the galaxy to its
spectroscopic redshift ($z=2.685$) measured from CO observations. The
best-fit SED and photometry are presented in
Figure~\ref{fig:Fig7} with the best-fit measured physical parameters
reported in Table 2. For star-formation rate, we used a Kennicutt
relation \citep{Kennicutt1998} with Chabrier initial mass function
\citep{Chabrier2003} to convert the total
infrared luminosities to SFR (${\rm SFR}=1\times10^{-10}L_{\rm IR}$;
\citealp{Riechers2013}). The SED of NA.v1.489 is consistent with the presence of a
Balmer break at $z\sim2.6$ and would allow for robust
stellar mass estimates.

It is important to note that the SED inferred parameters are for
stellar light heating the dust and does not include AGN
contribution. NA.v1.489 shows excess radio flux with respect to the
infrared luminosity compared to the average values expected for
high-$z$ SMGs which could hint towards the presence of an
AGN. Although there are studies such as \citet{Wang2013}, which show a
low fraction of AGN presence within SMG populations, there are studies
that point to the existence of an AGN component within SMGs \citep{Wang2013}.

The lack of an AGN recipe in our SED fitting with {\sc magphys} should not strongly
affect the estimated physical parameters \citep{Cunha2015}. In a recent study, in fact, using
synthetic models of galaxies with {\it H}-ATLAS like photometry
computed from simulations \citet{Hayward2015} showed that {\sc magphys}
estimated physical properties are robust even in the extreme case of
AGN contributing as much as 25\% to UV to IR luminosity.

Physical properties measured from SED fits have inherent
  uncertainties associated with choices of the input parameters in
  building the templates such as the assumed star-formation history
  and dust attenuation \citep{Conroy2013}. To further examine the robustness of our estimated physical
  properties, and in particular the stellar mass, we applied the SED fitting method detailed in
  \citet{michalowski08, michalowski09, michalowski10smg,
    michalowski10smg4, michalowski12mass, michalowski14mass}
  which is based on 35\,000 templates from the library of
\citet{iglesias07} plus some templates of \citet{silva98} and
\citet{michalowski08}, all of which were developed using
{\sc grasil}\footnote{\url{http://adlibitum.oats.inaf.it/silva/grasil/grasil.html}}
\citep{silva98}. They are based on numerical calculations of radiative
transfer within a galaxy, which is assumed to be a triaxial
axisymmetric system with diffuse dust and dense molecular clouds, in
which stars are born. The templates cover a broad range of galaxy properties from quiescent
to starburst and span an $A_V$ range from $0$ to $5.5$ mag. The
extinction curve (Figure 3 of \citealp{silva98}) is derived from the
modified dust grain size distribution of \citet{draine84}.
The star formation histories are assumed to be a smooth Schmidt-type
law i.e., the SFR is proportional to the gas mass to some
power; see \citep{silva98} with a starburst (if any) on top of
that, starting 50\,Myr before the time at which the SED is computed.
There are seven free parameters in the library of \citet{iglesias07}:
the normalization of the Schmidt-type law, the timescale of the mass
in-fall, the intensity of the starburst, the timescale for molecular
cloud destruction, the optical depth of the molecular clouds, the age
of the galaxy and the inclination of the disk with respect to the
observer. \citet{Michalowski2012, michalowski14mass} found that the choice of star
formation history (SFH) assumed in the SED modeling can lead to a
systematic shift of stellar mass. However, this shift is at most a
factor of $\sim2$ for a single-burst SFH (i.e. all stars formed at the
same epoch), unrealistic for such massive and actively star-forming
galaxy. Moreover, double-component SFH (utilized here) were found to
result in the most accurate stellar masses of real
\citep{Michalowski2012} and simulated SMGs
\citep{michalowski14mass, Hayward15}. We re-fit our observed SED with
the new set of templates using {\sc grasil} as discussed
above. Figure~\ref{fig:Fig7} shows the {\sc grasil} best-fit
template compared to the {\sc magphys}. Our newly measured physical
properties, and stellar mass, from the new templates are consistent
with our original measurements using {\sc magphys}.

\begin{table}
\begin{center}
\caption{Observed photometric data for NA.v1.489 (RA$^a$:13$^{\rm
  h}$35$^{\rm m}$42$^{\rm s}$.8,
Dec$^a$:+30$^{\circ}$03$^{\prime}$58$^{\prime\prime}$.1) at $z=2.685$.}
\begin{tabular}{lc}
\hline
\hline
Instrument & Flux Density \\ 
\hline
{\it HST}/WFC3 F110W & $1.16\pm0.39$ $\mu$Jy \\
Keck $H$ & $7.30\pm2.37$ $\mu$Jy \\ 
Keck $K_s$ & $18.06\pm4.07$ $\mu$Jy \\
{\it Spitzer}/IRAC 3.6\,$\mu$m & $67.18\pm7.10$ $\mu$Jy \\
{\it Spitzer}/IRAC 4.5\,$\mu$m & $93.36\pm5.41$ $\mu$Jy \\ 
WISE W3 & $115.0\pm16.0$ $\mu$Jy \\
WISE W4 & $911.3\pm190.4$ $\mu$Jy \\
{\it Herschel}/SPIRE 250\,$\mu$m & $136.6\pm7.2$ mJy \\
{\it Herschel}/SPIRE 350\,$\mu$m & $145.7\pm8.0$ mJy \\
{\it Herschel}/SPIRE 500\,$\mu$m & $125.0\pm8.5$ mJy \\
SMA 1.31\,mm & $5.93\pm0.43$ mJy \\
${\rm CO(3\rightarrow2)}$ 3.2\,mm Cont. & $874.6\pm55$ $\mu$Jy \\
JVLA 1.4\,GHz$^b$ & $134.9\pm21.8$ $\mu$Jy \\
\hline
\end{tabular}
\end{center}
\footnotesize
$^a$: From peak SMA. $^b$: From JVLA 6.89\,GHz observations assuming a
spectral index of $\alpha=-0.8$.
\end{table}

\begin{table}
\begin{center}
\caption{Measured physical properties of NA.v1.489.}
\begin{tabular}{ccc}
\hline
\hline
Quantity & Value & Unit \\ 
\hline
$M_{\star}$ & $6.8_{-2.7}^{+0.9}\times10^{11}$ & $M_{\odot}$ \\
$L_{\rm IR}$ & $1.9\pm0.2\times10^{13}$ & $L_{\odot}$ \\
$T_{\rm d}$ & $40\pm1$ & K\\
$M_{\rm d}$ & $1.5\pm0.3\times10^9$ & $M_{\odot}$\\
SFR$^a$ & $1914\pm180$ & $M_{\odot}{\rm yr^{-1}}$ \\
$L^{\prime b}_{\rm CO}$ & $1.0\pm0.1\times10^{11}$ & ${\rm K\,km\,s^{-1}\,pc^2}$ \\
$M^{b}_{\rm gas}$ & $8.3\pm1.0\times10^{10}$ & $M_{\odot}$\\
$L_{\rm IR}/L^{\prime}_{\rm CO}$ & $184$ & $L_{\odot}\,{\rm (K\,km\,s^{-1}\,pc^2)^{-1}}$\\
\hline
\end{tabular}
\end{center}
\footnotesize
$^a$: Assuming a Chabrier initial mass function and a conversion of
${\rm SFR}[M_{\odot}{\rm yr^{-1}}]=1.0\times10^{-10}\,L_{\rm
  IR}[L_{\odot}]$ \citep{Riechers2013}, $^b$: From GBT ${\rm
  CO(1\rightarrow0)}$ observations corrected for magnification
assuming $L_{\rm CO}^{\prime}-{\rm to}-M({\rm H_2})$ conversion factor
of $\alpha_{\rm CO} = 0.8\,M_{\odot}{\rm (K\,km\,s^{-1}\,pc^2)^{-1}}$. 
\end{table}

\subsection{Molecular Gas and Dust}

We use the velocity integrated CO flux ($S_{\rm CO}\,\Delta v$), measured from GBT
observations (Figure~\ref{fig:Fig2}), to estimate the CO line luminosity given the
spectroscopic redshift of the SMG. The best-fit Gaussian to the $\rm
{CO (1 \rightarrow 0)}$ at 31.28\,GHz yields a peak observed flux of 1.54\,mJy and
velocity FWHM of $305\pm87\,{\rm km/s}$. The CO line luminosities could be
calculated as \citep{Solomon2005, Ivison2011, Bolatto2013, Carilli2013, Scoville2016}:

\begin{equation}
\begin{split}
L_{\rm CO}^{\prime}[{\rm
  K\,km\,s^{-1}\,pc^2}]=3.25\times10^7(S\Delta\nu [{\rm
  Jy\,km\,s^{-1}}])\\
\times(\nu_{\rm obs}[{\rm GHz}])^{-2}(D_{\rm L} [{\rm Mpc}])^2(1+z)^{-3}
\end{split}
\end{equation}

This yields a magnification corrected ${\rm CO(1\rightarrow0)}$ line
luminosity of $L_{\rm
  CO(1\rightarrow0)}^{\prime}=1.04\pm0.12\times10^{11}\,{\rm
  K\,km\,s^{-1}\,pc^2}$. Assuming $L_{\rm CO}^{\prime}-{\rm to}-M({\rm
  H_2})$ conversion factor of $\alpha_{\rm CO} = 0.8\,M_{\odot}{\rm
  (K\,km\,s^{-1}\,pc^2)^{-1}}$, which is commonly adopted for
star-bursting galaxies \citep{Downes1998, Solomon2005,
  Tacconi2008}, we measure a molecular gas mass ($M({\rm
  H_2})=\alpha_{\rm CO}L^{\prime}_{\rm CO}$) of
$8.32\pm0.09\times10^{10}\,M_{\odot}$ for NA.v1.489.

\begin{figure}
\begin{center}
\includegraphics[trim=2cm 0cm 0cm 0cm, scale=0.48]{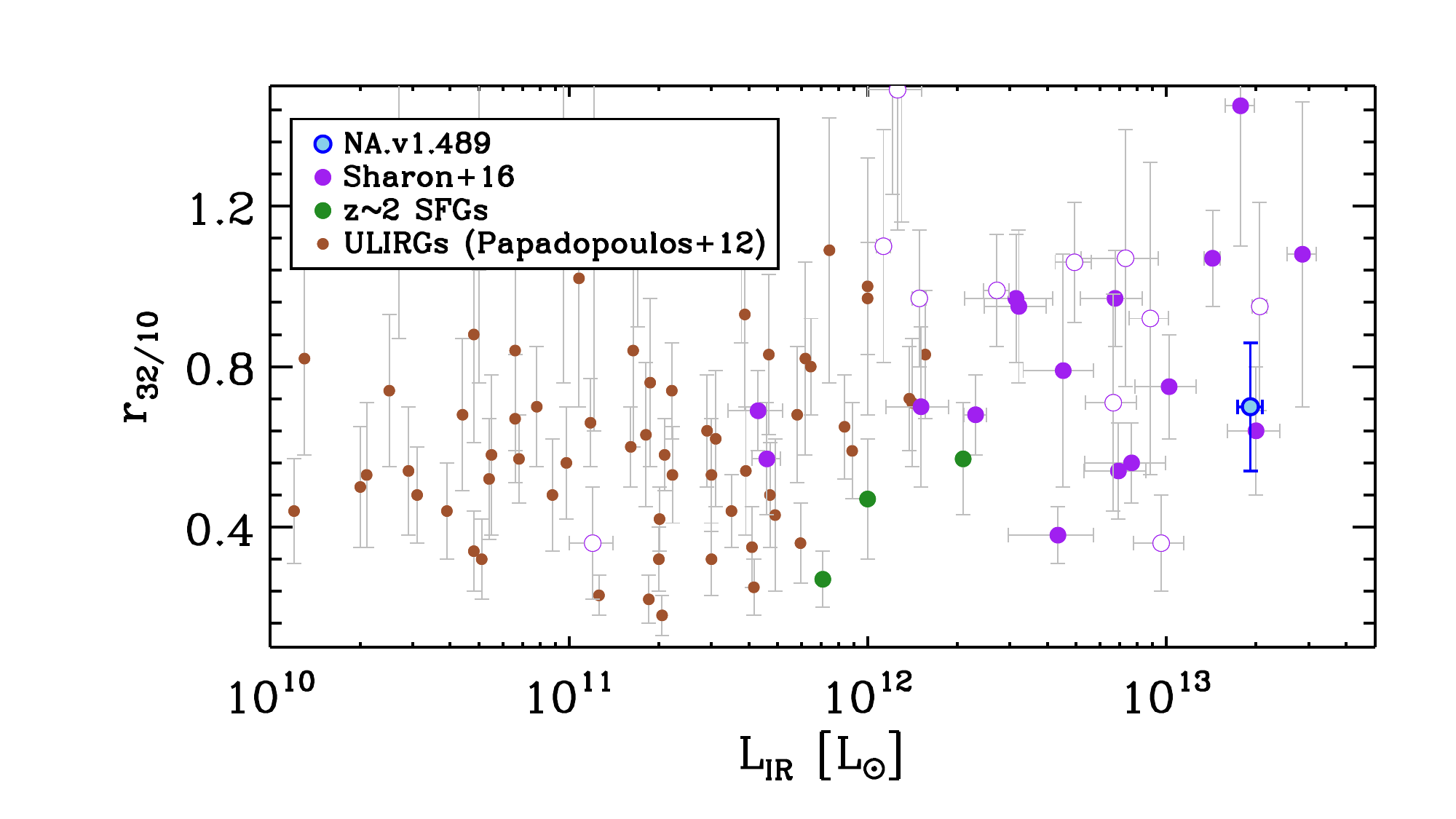} 
\caption{The molecular line intensity ratio of NA.v1.489 ($r_{32/10}=L_{\rm CO(3 \rightarrow2)}^{\prime}/L_{\rm CO(1\rightarrow0)}^{\prime}$) as measured by GBT and
  CARMA. The plot shows the same relation for local
  ULIRGs from \citet{Papadopoulos2012} (in brown), $z\sim2$ SMGs and quasars
  \citep{Sharon2016} (in magenta) and massive star forming galaxies
  \citep{Aravena2014, Daddi2015} (in green). The $z\sim2$ SMGs (filled magenta)
  have average line ratio of $r_{32/10}=0.78$ while the AGN host galaxies (open magenta)
  have average measured line ratios of $r_{32/10}=1.03$
  \citep{Sharon2016} where the $J_{\rm upper}=3$ line flux is boosted by
  emission from the AGN.}
\label{fig:Fig8}
\end{center}
\end{figure}

The $\rm {CO (3\rightarrow 2)}$ observations obtained with CARMA yield
a magnification corrected line
luminosity of $L_{\rm CO(3\rightarrow2)}^{\prime}=7.30\pm0.12\times10^{10}\,{\rm
  K\,km\,s^{-1}\,pc^2}$. This yields a line
ratio of $r_{32/10}=L_{\rm CO(3 \rightarrow2)}^{\prime}/L_{\rm CO(1
  \rightarrow0)}^{\prime}=0.70\pm0.16$. This is similar to the
  line ratios derived from large samples of SMGs at $z\sim2$ with
  $r_{32/10}=0.78\pm0.27$ \citep{Sharon2016}.

The ISM mass can also be estimated from sub-mm
continuum emission \citep{Magdis2012, Scoville2014, Scoville2016} as
the dust emission is optically thin at the long
wavelength. \citet{Scoville2016} provide an empirical calibration to measure
the ISM from the Rayleigh-Jeans tail emission at 850\,$\mu$m:

\begin{equation}
\begin{split}
M_{\rm ISM}&=1.78 \times S^{\rm obs}_{\nu} [{\rm mJy}] \times
(1+z)^{-4.8}\\
&\times \bigg(\frac{\nu_{850}}{\nu_{\rm obs}}\bigg)^{3.8}(D_{\rm
  L}[{\rm Gpc}])^2\\
&\times
\bigg\{\frac{6.7\times10^{19}}{\alpha_{850}}\bigg\}\frac{\Gamma_0}{\Gamma_{\rm
  RJ}}\times10^{10}\,M_{\odot}
\end{split}
\end{equation}

where $\Gamma_{\rm RJ}$ is the correction applied for deviations from
the $\nu^2$ behavior in the RJ tail as defined in \citet{Scoville2014}
and \citet{Scoville2016}. We used the SMA observations at 1.31\,mm and
measured a magnification corrected ISM mass of
$7.92\times10^{10}\,M_{\odot}$. This is consistent with the molecular
gas mass estimates derived independently from GBT. From CO
observations, we measure a gas depletion time scale ($t_{\rm
  dep}\equiv M_{\rm gas}/{\rm SFR}$)) of 43\,Myr, indicating a rapid phase
of star-formation which is also observed in other high redshift SMGs
and starbursts \citep{Tacconi2008, Messias2014,
  Oteo2016}. 

\begin{figure*}
\centering
\leavevmode
\includegraphics[trim=2cm 0cm 0cm 0cm, scale=0.47]{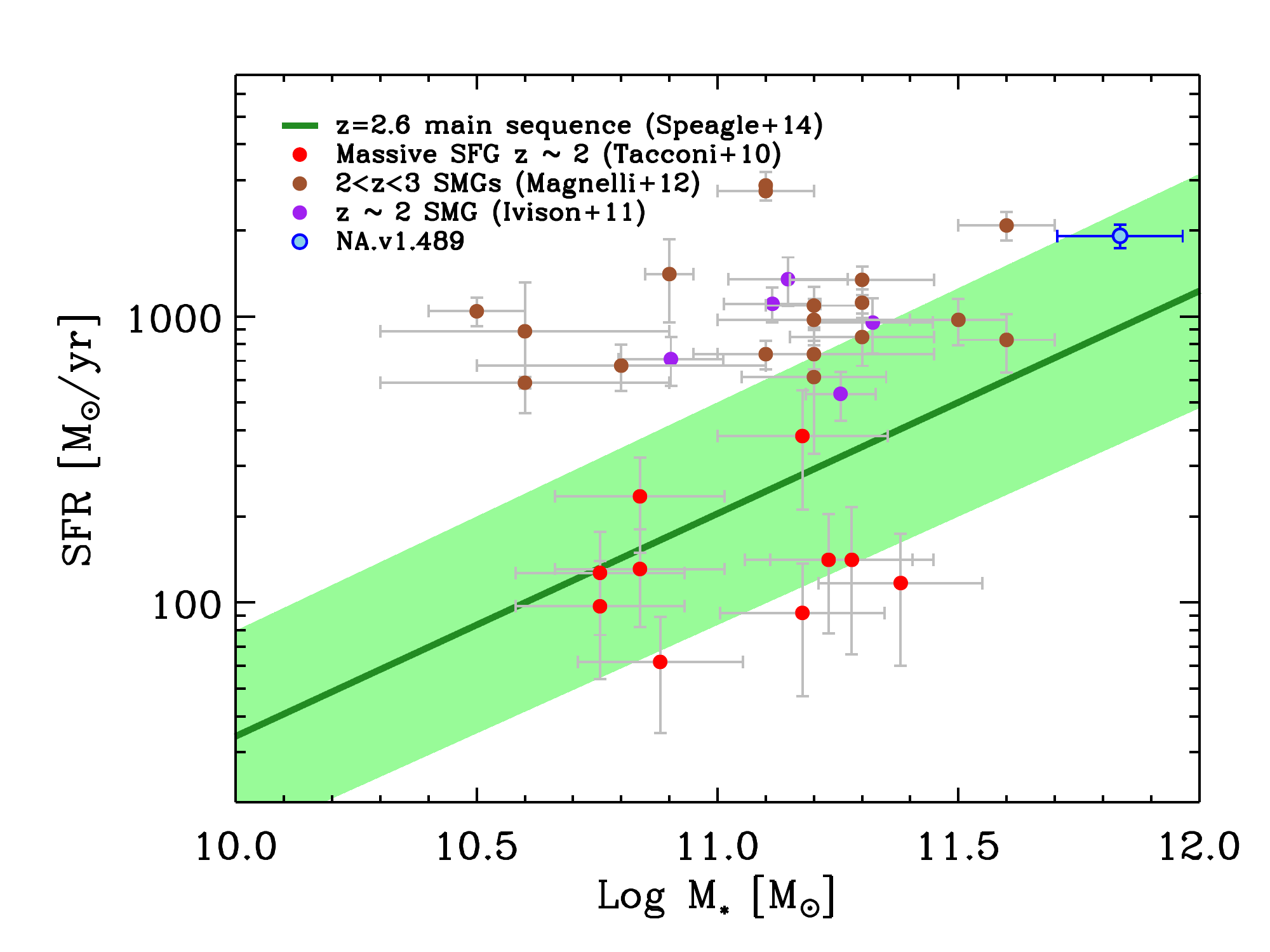} 
\includegraphics[trim=2cm 0cm 0cm 0cm, scale=0.47]{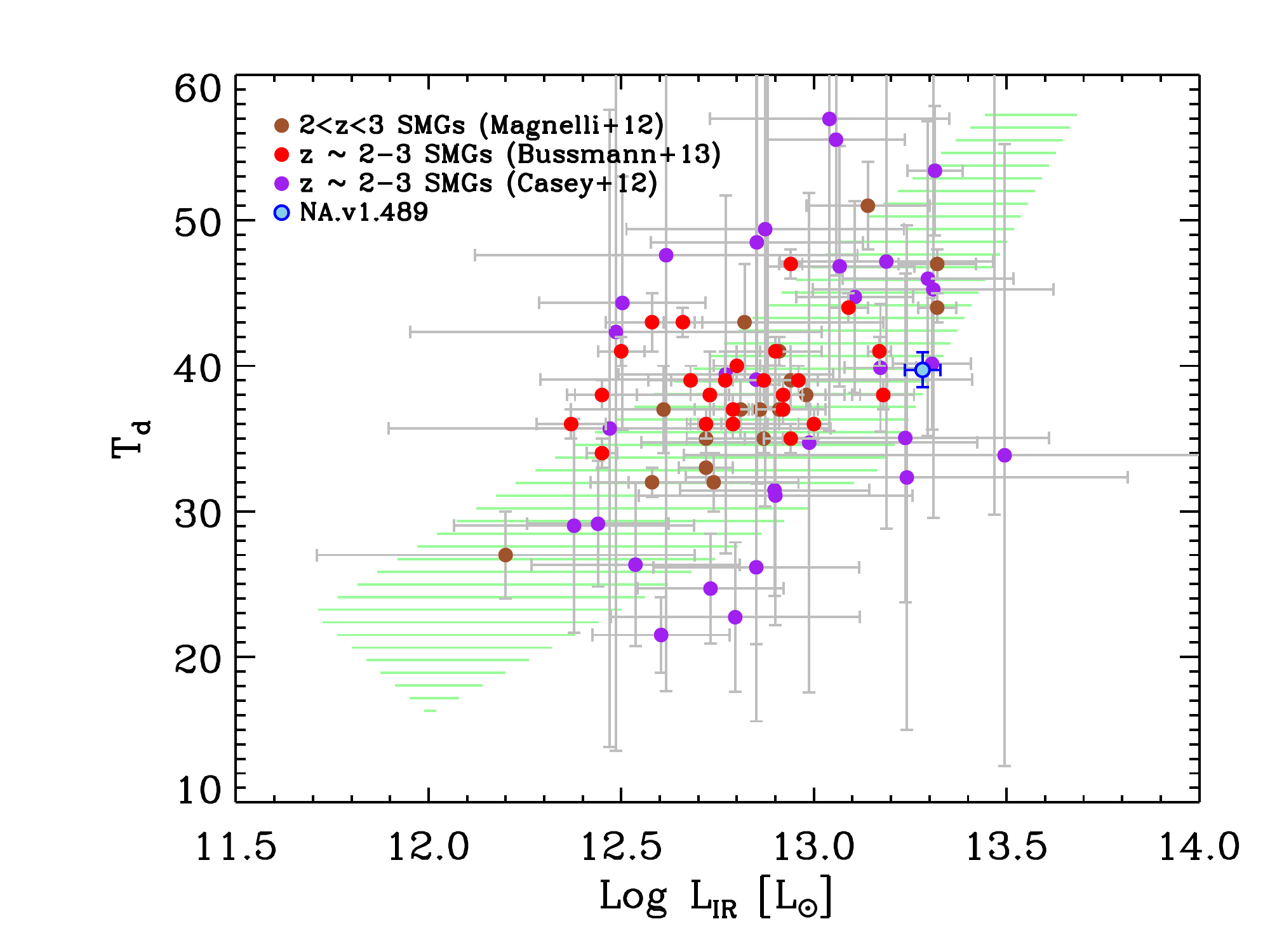}
\caption{{\it Left}: The main sequence of star formation. The green line shows
  the expected trend for $z=2.6$ star forming galaxies
  reported by \citet {Speagle2014}. NA.v1.489 is marked with the
blue circle as a massive SMG at $z\sim2.6$ along with SMGs
\citep{Ivison2011, Magnelli2012}
and massive star-forming galaxies \citep{Tacconi2010} at
similar redshifts. {\it Right}: The dust temperature and bolometric
infrared luminosity of sub-millimeter galaxies \citep{Magnelli2012, Casey2012, Bussmann2013}. The
green shaded area is the measured locus of SMGs from
\citet{Chapman2005}. The scatter around
the relation is associated with selection biases \citep{Wardlow2011,
  Magnelli2012}. NA.v1.489 has colder dust temperature compared to
SMGs at similar redshift, disfavoring a major merger scenario.}
\label{fig:Fig9}
\end{figure*}

\begin{figure*}
\centering
\includegraphics[trim=2cm 0cm 0cm 0cm, scale=0.32]{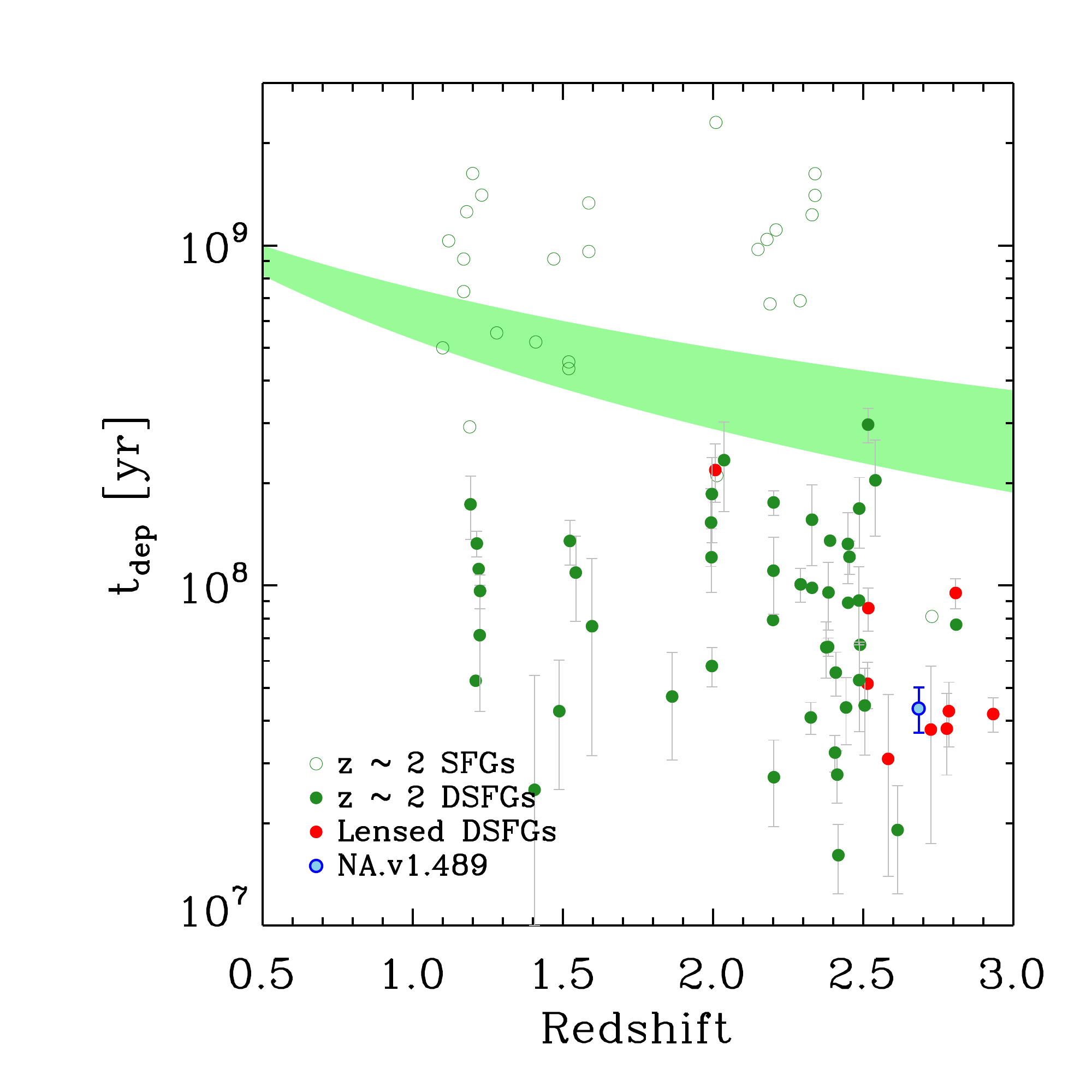} 
\includegraphics[trim=2cm 0cm 0cm 0cm, scale=0.32]{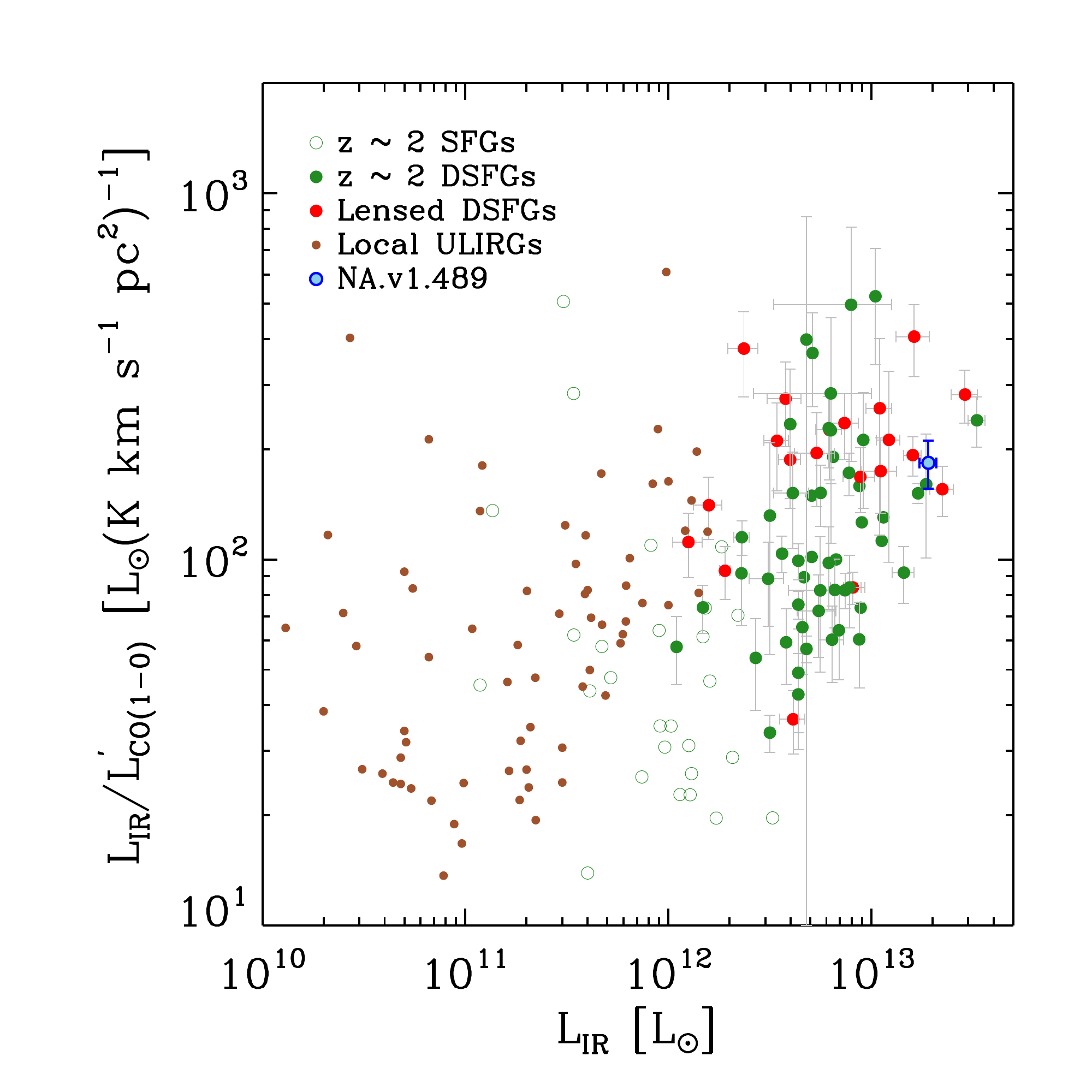}
\includegraphics[trim=2cm 0cm 0cm 0cm, scale=0.32]{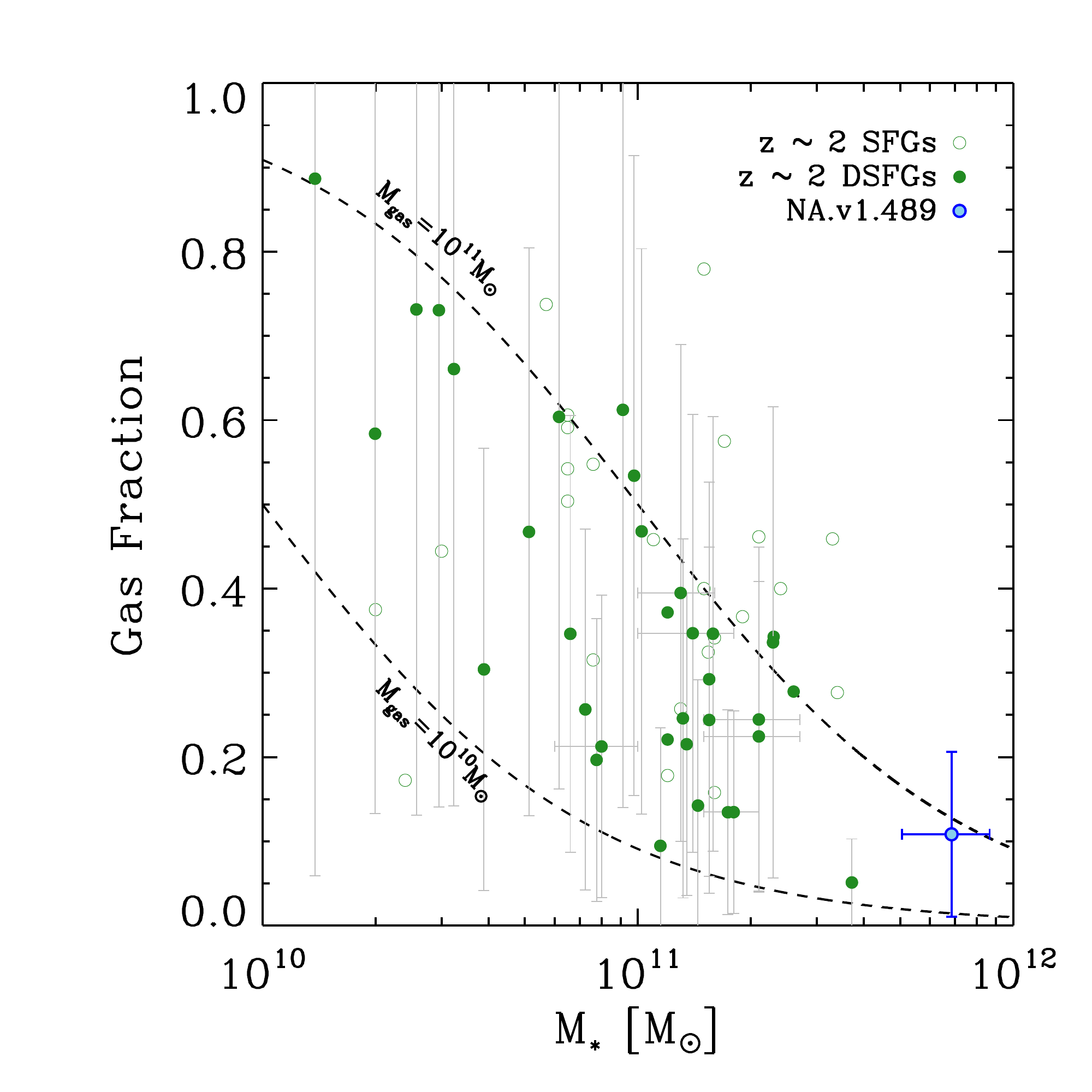}
\caption{{\it Left}: The evolution of the gas depletion time scale
  ($t_{\rm dep}\equiv M_{\rm gas}[M_{\odot}]/{\rm SFR}[M_{\odot}{\rm yr^{-1}}]$)
  out to $z\sim3$. The open symbols show the gas depletion for normal
  and lensed star-forming galaxies \citep{Genzel2010, Dessauges2015} compared to the shorter
  time scales seen for SMGs (filled green symbols;
  \citealp{Frayer2008, Ivison2010, Ivison2011, Ivison2013, Bothwell2013}). The
  shaded area shows the predicted evolution of main sequence galaxies
  at $z<3$ \citep{Dave2012, Saintonge2013}. NA.v1.489 has a gas depletion
  time scale of 43\,Myr which is much lower than the SFGs at similar
  redshifts and consistent with lensed SMGs identified with the South
Pole Telescope (SPT; \citealp{Aravena2016}) and from
\citet{Thomson2012} at similar redshifts. {\it Middle}: The
star-formation efficiency (SFE) vs. infrared luminosity ($ L_{\rm
  IR}$; rest-frame $\rm 8-1000\,\mu m$) of NA.v1.489 compared to local
ULIRGs \citep{Papadopoulos2012} and high redshift star-forming and
sub-millimeter galaxies. {\it Right}: The molecular gas fraction ($f_{\rm
  g}\equiv M_{\rm g}/M_{\star}+M_{\rm g}$) as a function of stellar
mass. NA.v1.489 has a low gas fraction compared to SMGs at similar
redshifts which could be indicating that the galaxy is towards the end
of the star-formation phase with near complete mass build-up. The
dashed lines show tracks for constant gas masses of
$10^{10}\,M_{\odot}$ and $10^{11}\,M_{\odot}$.}
\label{fig:Fig10}
\end{figure*}

\section{Discussion}

NA.v1.489 has a molecular gas line ratio of $r_{32/10}=0.70\pm0.16$ as
measured from GBT and CARMA. This is higher than the measured
  ratio for normal star-forming galaxies \citep{Aravena2014, Daddi2015},
  with $r_{32/10}=0.42\pm0.07$ \citep{Daddi2015}. The larger value is associated with the
increased star-formation activity which could raise the higher-$J$ CO
line fluxes at higher energies \citep{Bolatto2013}. The measured molecular gas line
flux ratio of NA.v1.489 further agrees with the mean
value reported by \citet{Sharon2016} for SMGs ($0.78\pm0.27$) at
$z\sim2$. Figure~\ref{fig:Fig8} shows
the CO molecular line flux ratios as a function of infrared luminosity
for NA.v1.489 compared to other SMGs and QSO host galaxies at $z\sim2$
\citep{Sharon2016}, normal star-forming galaxies \citep{Aravena2014,
  Daddi2015} and local ULIRGs \citep{Papadopoulos2012}. Although the
excitation temperature of $\rm {CO (3\rightarrow2)}$ is not very high
and could well be produced by a compact starburst \citep{Sharon2016}, a particularly
energetic source, such as an AGN, could further increase the
$r_{32/10}$ line ratios. NA.v1.489 has measured line ratios consistent
with the line excitation produced by a starburst rather than a QSO host
galaxy (which has $r_{32/10}=1.03\pm0.50$; \citealp{Sharon2016}). However,
given the uncertainties in both measurements, line excitation due to the
presence of an AGN cannot be ruled out.

NA.v1.489 has an SED inferred infrared luminosity that puts it among the
most IR-luminous extra-galactic sources and which then yields an SED inferred SFR
comparable to the most intense star forming systems at $z \sim 2-3$
\citep {Greve2005, Harris2012, Magnelli2012, Fu2012, Fu2013, Ivison2013}. Figure~\ref{fig:Fig9}
shows the SFR versus stellar mass plot along with the main sequence of
star forming galaxies \citep{Kauffmann2003, Noeske2007,
  Elbaz2011, Shivaei2015} derived from \citet{Speagle2014}. The $z=2.6$ main
sequence relation sits above the corresponding local relation such
that at any given fixed stellar mass the high-$z$ counter part would
be much more star forming compared to the local SFGs. The scatter along
this relation, marked as the green shaded area, is associated to the
intrinsic properties of galaxies such as the change in star formation histories and metallicities
\citep {Brinchmann2004, Mannucci2010, Wuyts2011}. As we see from
Figure~\ref{fig:Fig9}, NA.v1.489 sits at the
main sequence relation plotted for $z=2.6$ from \citet
{Speagle2014}. This is supported by recent studies of high redshift
SMGs pointing towards small deviations from the SFG main sequence
\citep{Michalowski2012, Koprowski2014, Koprowski2016}. The intense SFR agrees with the MS
position given the large stellar mass of the system and the huge
reservoirs of molecular gas available. The large stellar mass is in
agreement with simulations of $z\sim2$ SMGs \citep{Dave2010} and SMG
models of \citet{Hayward2011} requiring large stellar masses for 
bright flux densities in the sub-mm (with $M_{\star}>6\times10^{10}$
needed to create a 850\,$\mu$m flux density of 3\,mJy) also seen in
observed stellar mass estimates of SMGs at
$z\sim2$ from multi-wavelength SED measurements
\citep{Michalowski2010, Hainline2011}. The
intense SFR could be explained as being due the
higher molecular gas mass available within these systems that provide
the material needed for the excess star formation \citep
{Riechers2010, Fu2013, Riechers2014}. The dust temperature of SMGs
derived from far-IR observations is
indicative of the ISM in these systems. The local infra-red
  luminous sub-millimeter galaxies show a clear trend between the dust
temperature ($T_{\rm d}$) and the infrared luminosity \citep{Hwang2010 ,
  Elbaz2011, Symeonidis2013} with similar trends observed for high redshift
far-infrared or sub-millimeter selected samples of galaxies
\citep{Chapman2005, Magdis2010, Magnelli2012, Casey2012, Magnelli2014,
  Bethermin2015}. At high-redshift, these
relations are biased towards colder temperatures for sub-mm selected
samples at low $L_{\rm IR}$ \citep{Casey2009, Magnelli2012} as direct observations in the
far-infrared are mostly limited to bright sources at $z\sim2$
\citep{Magdis2010} with studies of far-infrared lensed systems extending this to
lower $L_{\rm IR}$ \citep{Bussmann2013, Nayyeri2016}. The $T_{\rm d}-L_{\rm IR}$
relation shows modest evolution with redshift from {\it Herschel}-selected
samples out to $z\sim2-3$ \citep{Hwang2010} where SMGs at high-$z$
are found to be on average $\rm 2-5\,K$ colder than the local
counterparts \citep{Hwang2010}. Figure~\ref{fig:Fig9}
shows that NA.v1.489 has dust temperature ($T_{\rm d}=40\,{\rm K}$) and far-IR
luminosity ($L_{\rm IR}=1.9\times10^{13}\,L_{\odot}$) that
agrees with the general trend observed for the SMGs
\citep{Chapman2005}. The scatter in this relation is associated with
selection biases as well as the presence of colder dust temperatures
in high-$z$ SMGs (compared to the local counterparts)
\citep{Magnelli2012}. Figure~\ref{fig:Fig9} further shows
that NA.v1.489 on average has a colder dust temperature compared to sub-mm and
{\it Herschel}-selected samples of SMGs at $z\sim2-3$ at similar infrared luminosities
\citep{Casey2012, Magnelli2012}. The lower dust temperature measured for
NA.v1.489 disfavors a major merger scenario. Furthermore the colder $T_{\rm
  d}$ is consistent with a more extended dust distribution
\citep{Hwang2010} as observed from our SMA observations.

The large molecular gas measured from CO observations is responsible
for the intense star-formation observed in NA.v1.489. Similar
reservoirs of cold molecular gas are observed in other extreme
starbursts at high redshift \citep{Riechers2011, Riechers2013, Fu2013,
Spilker2015}. NA.v1.489 has a gas depletion
time-scale of ($t_{\rm dep}\equiv M_{gas}/{\rm SFR}$) 43\,Myr, much
shorter than the star-forming galaxies $\rm \sim1\,Gyr$ at similar
redshifts \citep{Genzel2010, Decarli2016a, Decarli2016b} and
  comparable to gas depletion time-scales in other sub-mm selected
  dusty galaxies at high redshift \citep{Riechers2010, Riechers2013}. The short time-scale
and intense star-formation rates are responsible for the mass build-up
in NA.v1.489. Figure~\ref{fig:Fig10} shows the gas depletion
time-scale of NA.v1.489 compared to normal star-forming and
sub-millimeter galaxies at similar and lower redshifts. The star-formation rate
efficiency of NA.v1.489 (${\rm SFE}\equiv L_{\rm IR}/L_{\rm
  CO}^{\prime}$) is much higher than normal star forming galaxies. The
intense star-formation in NA.v1.489 is not only because of the huge
molecular gas reservoirs, which is comparable in some cases to those
of massive star-forming galaxies \citep{Genzel2010, Tacconi2013,
  Genzel2015}, but also to higher SFE that is also observed in other
high-$z$ SMGs \citep{Fu2013}. Figure~\ref{fig:Fig10} shows the SFE of
NA.v1.489 as a function of infrared luminosity ($ L_{\rm IR}$;
rest-frame $\rm 8-1000\,\mu m$) compared to the normal star-forming
and sub-millimeter galaxies. NA.v1.489 in particular has a higher SFE
compared to normal star-forming galaxies and local ULIRGs which is
also observed in other SMGs at similar redshift
\citep{Fu2012}. NA.v1.489, in particular, has a low gas fraction (with
$f_{\rm g}=11\%$) compared to less massive normal star-forming and
sub-millimeter galaxies (Figure~\ref{fig:Fig10}). Given the
high stellar mass and low gas fraction, NA.v1.489 is a SMG that
likely has already formed most of the stars.  

The molecular gas mass as measured by the CO line luminosity depends
on the assumed value for the conversion factor ($\alpha_{\rm
  CO}$). One of the main uncertainties in determining the total gas
mass in fact lies with the uncertainties associated with the ${\rm
  CO-H_2}$ conversion factor $\alpha_{\rm CO}$ \citep {Narayanan2012,
  Spilker2015}. In this work, we assumed a conversion factor of
$\alpha_{\rm CO} = 0.8\,M_{\odot}{\rm (K\,km\,s^{-1}\,pc^2)^{-1}}$ to
estimate the total molecular gas mass from CO observations. This is
lower than the assumed value for the Milky Way ($\alpha_{\rm CO,MW} = 4.36\,M_{\odot}{\rm
  (K\,km\,s^{-1}\,pc^2)^{-1}}$ measured from direct observations of
molecular Hydrogen and CO; \citealp{Strong1996, Leroy2011, Bolatto2013}) and that of nearby and
high redshift normal star-forming galaxies \citep{Sandstrom2013,
  Tacconi2013}. The lower conversion factor is more favored for
the extreme starburst due to higher gas temperature and velocity
dispersions \citep{Narayanan2011, Narayanan2012, Papadopoulos2012, Bolatto2013} with the
higher values overestimating the molecular gas mass in these systems
\citep{Downes1993, Solomon1997}. This is supported by the dust temperature measured
for NA.v1.489 which could be a proxy for gas temperature
\citep{Yao2003}. For normal star-forming galaxies
analogs of NA.v1.489 (i.e. at similar stellar mass and
redshift) $\alpha_{\rm CO}$ could be measured given the metallicity dependent relation provided by
\citet{Genzel2015} (see Equations $6-8$ in that paper; see also
\citealp{Dessauges2016}), where metallicity is estimated at the redshift and
stellar mass of NA.v1.489 following the mass-metallicity relation
(\citealp{Erb2006, Maiolino2008, Wuyts2014, Zahid2014}; see
Equation 12 in \citet{Genzel2015}). This yields a higher $\alpha_{\rm
  CO}$ compared to what is assumed for starbursts overestimating the
molecular gas. Given the uncertainties one could
derive a possible range of molecular gas masses following
\citet{Ivison2011} for a range of possible
conversion factors. Here we report the gas mass estimate based on
$\alpha_{\rm CO} = 0.8\,M_{\odot}{\rm (K\,km\,s^{-1}\,pc^2)^{-1}}$
which is consistent with what was discussed above.

We investigate the far infrared and radio properties of NA.v1.489 by
looking at the FIR/radio ratio ($q_{\rm IR} \equiv {\rm
  log_{10}}(S_{\rm IR}/3.75\times10^{12}\,{\rm W\,m^{-2}})-{\rm
  log_{10}}(S_{\rm 1.4\,GHz}/{\rm W\,m^{-2}\,Hz^{-1}})$) and how it compares with the
SMGs at similar and lower redshifts. We measured a $q_{\rm IR}=3.22$ for NA.v1.489
which is consistent with the average value derived for samples of
far-infrared selected galaxies at lower redshifts \citep{Ivison2010b} the existence of
which reflects the correlation between the processes that produce the
underlying emissions; namely star formation activity and synchrotron
radiation from supernova processes contributing to FIR and radio emission
respectively \citep{Yun2001, Bell2003, Chapman2005, Ivison2010b,
  Elbaz2011, Pannella2015}. The measured $q_{\rm IR}$ is consistent
with low redshift \citep{Ivison2010b, Michalowski2010} and high redshift estimates
\citep{Pannella2015} as well as the average value for SMGs
\citep{Ivison2010b, Pannella2015}. In fact \citet{Ivison2010b} showed
that the ratio is not evolving with time. The measured radio
luminosity of NA.v1.489 ($L_{1.4}=4.3\times10^{24}\,{\rm W\,Hz^{-1}}$) and FIR/radio ratio indicate
that NA.v1.489 is not dominated by a radio-loud AGN (with
$L_{1.4}\geq10^{25}\,{\rm W\,Hz^{-1}}$) or having a radio excess (with
$q_{\rm IR}\leq1.64$) as defined by \citet{Yun2001}. The radio
luminosity ($\rm L_{1.4\,GHz}$) as measured by the JVLA,
yields a ${\rm SFR} \sim 2560\pm414\,M_{\odot}{\rm yr^{-1}}$ using the calibration in
\citet{Pannella2015}. This agrees with the SFR measured
previously from the infrared luminosity derived from multi-band SED
fit with the difference associated with the uncertainties that exist in the
calibrations.

\section{Summary and Conclusion}

We have presented a detailed study of a massive {\it
    Herschel}/SPIRE detected sub-millimeter galaxy at $z=2.685$,
  gravitationally lensed by a {\it Spitzer} and WISE detected
  cluster at $z\sim1$. We provide detailed lens modeling of the system through combined Keck and Gemini spectroscopic and imaging
  observations used to identify foreground cluster members and lensing
  multiple images. Our best lens model provide a lensing
  magnification of $\mu_{\rm star}=2.10$ and $\mu_{\rm dust}=2.02$ for the stellar and dust
  emissions respectively for NA.v1.489. Multi-band SED fitting, corrected for the lensing
  magnification, provides physical properties of NA.v1.489 including
  stellar mass ($M_{\star}$), total infrared luminosity ($L_{\rm IR}$;
  rest-frame $8-1000\,\mu$m), dust mass ($M_{\rm d}$) and dust
  temperature ($T_{\rm d}$); see Table 2. We further derived total
  molecular gas and explored ionization state of the gas from
  spectroscopic observations of the CO molecular emission lines. The
  combined physical properties derived from the SED fits and
  spectroscopic molecular CO line observations reveal that NA.v1.489 is a
  very massive sub-millimeter galaxy with ${\rm
    SFR}\sim2000\,M_{\odot}{\rm yr^{-1}}$, putting it as one of the most
  extreme starbursts during peak epoch of star-formation, without a
  dominant AGN mode. The
  low-$J$ molecular line luminosities revealed the existence of large
  reservoir of molecular gas that is being rapidly converted into
  stars at a much higher pace compared to that of local ULIRGs (see
  Figure \ref{fig:Fig10}). This combined with the measured low gas fraction ($f_{\rm g}=11\%$) and
  high stellar mass, shows that NA.v1.489 has already assembled most of
  its mass and is most likely on its way to becoming quiescent,
  forming a progenitor for the most massive early type galaxies
  found in the local Universe. 

Observations of this galaxy seem to suggest that the location of SMGs
on a ${\rm SFR}-M_{\star}$ diagram is consistent with the main
sequence of star-formation at the very massive end (Figure \ref{fig:Fig9};
\citealp{Speagle2014}) while still being above the relation which is supported by
excessive star-formation rate driving the intense infrared
luminosities. This is in contrast to less massive SMGs with much
higher observed SFRs compared to normal star-forming galaxies at
similar stellar masses at $z\sim2$ (Figure \ref{fig:Fig9};
\citealp{Tacconi2010,Magnelli2012}). As discussed above, this indicate that the very
massive SMGs are on their way to becoming quiescent systems. We note
here that in addition to uncertainties associated with stellar mass (which we addressed
in Section 4), uncertainties in the magnifications measured from the lens model could also
lead to a high stellar mass. To test the robustness of the derived
magnifications in the lens model, we recomputed the magnification
using MCMC maximum likelihood which yielded the original magnification factor.

One of the main contributing factors to the transition of galaxies
from a star-forming phase to that of quiescence is the feedback
associated with AGN activity \citep{Sijacki2007, Feruglio2010, Cicone2014} diluting and heating up the
molecular gas ceasing star-formation \citep{Scannapieco2005}. This generally begins with the
central black hole gaining mass through constant accretion after which
the AGN is ignited \citep{Hopkins2016}. The AGN signature has been observed
in SEDs of SMGs in the mid-infrared
and X-ray \citep{Alexander2005, Valiante2007, Laird2010}. The
far-infrared luminosities are however believed to be
mainly driven by star-formation activity rather than AGN
\citep{Magnelli2012}. High emission line excitation ratios could be evidence for the
presence of an AGN, though a lack of such highly excited lines does not
necessarily rule out the existence of it. A combined X-ray through
radio observations of distant SMGs along with mid-infrared
constrained SEDs and emission line ratios is needed to study the
presence and role of the AGNs in detail. As
discussed above, although we did not see a signature of a dominant
AGN mode from CO line ratios in NA.v1.489, the presence of such
component could not be ruled out with current observations.

\section*{Acknowledgement} 

We wish to thank the anonymous referee for reading the original
manuscript and providing useful suggestions. Financial support for this work was provided by NSF
through AST-1313319 for HN and AC. UCI group also acknowledges
support from HST-GO-14083.002-A, HST-GO-13718.002-A and NASA
NNX16AF38G grants. MN has
received funding from the European Unions Horizon 2020 research and
innovation program under the Marie Sklodowska-Curie grant agreement No
707601. DR acknowledges
 support from the National Science Foundation under grant number
 AST-1614213 to Cornell University. GDZ acknowledges
financial support by ASI/INAF agreement n. 2014-024-R.0.. HD
acknowledges financial support from the Spanish Ministry of Economy
and Competitiveness (MINECO) under the 2014 Ram\'{o}n y Cajal program
MINECO RYC-2014-15686. IO and RJI acknowledge support from the
European Research Council in the form of the Advanced Investigator
Programme, 321302, {\sc cosmicism}. JLW is supported by a European
Union COFUND/Durham Junior Research Fellowship under EU grant
agreement number 267209, and acknowledges additional support from STFC
(ST/L00075X/1). Some of
the data presented herein were obtained at the W.M. Keck Observatory,
which is operated as a scientific partnership among the California
Institute of Technology, the University of California and the National
Aeronautics and Space Administration. The Observatory was made
possible by the generous financial support of the W.M. Keck
Foundation. The authors wish to recognize and acknowledge the very
significant cultural role and reverence that the summit of Mauna Kea
has always had within the indigenous Hawaiian community. We are most
fortunate to have the opportunity to conduct observations from this
mountain. Data presented herein were obtained using the UCI Remote Observing Facility, 
made possible by a generous gift from John and Ruth Ann Evans. Support for CARMA
construction was derived from the Gordon and Betty Moore
Foundation, the Kenneth T. and Eileen L. Norris Foundation,
the James S. McDonnell Foundation, the Associates of the California
Institute of Technology, the University of Chicago, the
states of California, Illinois, and Maryland, and the National
Science Foundation. Ongoing CARMA development and operations
are supported by NSF grant ATI-0838178 to CARMA,
and by the CARMA partner universities. The Submillimeter Array is a
joint project between the Smithsonian Astrophysical Observatory and
the Academia Sinica Institute of Astronomy and Astrophysics and is
funded by the Smithsonian Institution and the Academia Sinica.

\bibliographystyle{apj}
\bibliography{references}

\end{document}